\documentclass[a4paper]{article}

\usepackage{INTERSPEECH2021}
\usepackage{caption}
\usepackage{stfloats}
\graphicspath{{./figures/}}
\DeclareGraphicsExtensions{.pdf,.jpg,.png}
\usepackage[hidelinks]{hyperref}       
\usepackage{url}            
\usepackage{balance}

\title{Evaluating Features and Metrics for High-Quality Simulation of \\
Early Vocal Learning of Vowels}
\name{Branislav Gerazov$^1$, Daniel van Niekerk$^2$, Anqi Xu$^2$, Paul Konstantin Krug$^3$, \\Peter Birkholz$^3$, and Yi Xu$^2$}
\address{
  $^1$Faculty of Electrical Engineering and Information Technologies, CMUS, Skopje, Macedonia\\
$^2$Department of Speech, Hearing and Phonetic Sciences, University College London , UK\\
$^3$Institute of Acoustics and Speech Communication, TU Dresden, Germany
}
\email{gerazov@feit.ukim.edu.mk}

\begin{document}

\maketitle
\begin{abstract}
 The way infants use auditory cues to learn to speak despite the acoustic mismatch of their vocal apparatus is a hot topic of scientific debate. The simulation of early vocal learning using articulatory speech synthesis offers a way towards gaining a deeper understanding of this process. One of the crucial parameters in these simulations is the choice of features and a metric to evaluate the acoustic error between the synthesised sound and the reference target. We contribute with evaluating the performance of a set of 40 feature-metric combinations for the task of optimising the production of static vowels with a high-quality articulatory synthesiser. Towards this end we assess the usability of formant error and the projection of the feature-metric error surface in the normalised F1-F2 formant space. We show that this approach can be used to evaluate the impact of features and metrics and also to offer insight to perceptual results.
\end{abstract}

\section{Introduction}

The way infants learn to speak is a hot topic of scientific debate.
The process is likely driven by the auditory perception of language in their surroundings \cite{kuhl2000anew, vihman1994nature}, which is reinforced by the fact that children born blind learn how to speak on their own \cite{perez2019language}, while those born deaf cannot \cite{oller1983development}.
Albeit, the absence of visual cues does hinder proper articulation of phonemes such as /u/, which has been found less rounded in the blind \cite{menard2013acoustic}.
Still, it is a mystery how infants use auditory cues to generate matching vocalisations in light of the differences in the size of their vocal apparatus \cite{fitch1999morphology, messum2015creating, breazeal2002robots}.

A crucial way towards gaining a deeper understanding of this process is through the simulation of early vocal learning based on articulatory speech synthesis \cite{howard2011modeling, rasilo2017online, prom2013training, prom2014identifying}.
In its basic form this approach relies on the optimisation of the parameters of the synthesiser, based on the acoustic comparison of the synthesised speech to a template \cite{prom2013training}, but some have used it as a part of more complex models of speech motor control \cite{parrell2019facts}.
Using such an experimental setup researchers have successfully simulated the need of visual cues of lip rounding to synthesise high quality rounded vowels \cite{murakami2015seeing}.
Others have used it to test hypotheses that the burden of speaker normalisation during vocal learning is on the adults \cite{howard2011modeling, messum2015creating}, but synthetic speech simulated using adult mimicry of babbles yielded low vowel identification scores \cite{rasilo2017online}.
Some have successfully simulated vocal learning of syllables \cite{prom2013training, xu2019coarticulation}.

One crucial part of these systems is the choice of features used to represent the speech signals and the distance metric used to compare them to determine the articulation error that drives learning.
Formant error has been used extensively for simulations of vowel learning \cite{parrell2019facts, rasilo2017online}, with another common approach being the use of auditory filterbanks \cite{howard2011modeling} and especially Mel-Frequency Cepstral Coefficients (MFCCs), perhaps owing to their predominance in Automatic Speech Recognition (ASR) \cite{young2006htk, povey2011kaldi}.
Prom-on et al.\ used the sum of squares MFCC error as a metric equivalent to the Mean Square Error (MSE) for optimisation \cite{prom2013training, prom2014identifying}.
Gao et al.\ used 13 MFCCs augmented by a probability of voicing and their 1st and 2nd derivatives in conjunction with the cosine distance \cite{gao2019articulatory}.
Other more advanced approaches have used models of peripheral processing of the cochlea and auditory memory \cite{murakami2015seeing}.

Despite of its importance this issue has not been analysed in detail and there is no consensus on which features and metric to use for simulating vocal learning, both based on their performance and on their physiological plausibility.
We  contribute here through the evaluation of 40 feature-metric pairs for the task of optimising the production of vowel targets with an articulatory synthesiser.
Specifically, our goal is to explore the impacts of:
\emph{i}) high frequency (HF) emphasis in the feature extraction process,
\emph{ii}) feature normalisation,
\emph{iii}) the use of different distance metrics, and
\emph{iv}) the use of different features.
Towards this end we assess the usability of two objective methods in this evaluation: the formant error of the optimised sounds in the normalised F1-F2 formant space, and the projection of the feature-metric's error surface in this space.
In addition, we explore if these methods can be used to augment or interpret perceptual scores.

\section{Methodology}

\subsection{Dataset}

\textbf{Vocal tract model.}
We used the VocalTractLab (VTL) API to synthesise the speech waveforms~\cite{birkholz2013modeling, birkholz2006construction}.\footnote{VTL v.2.2 \url{http://www.vocaltractlab.de/}}
VTL is an articulatory synthesizer that synthesises audio using acoustic simulations based on the crossarea of the vocal tract calculated from a geometrical 3D vocal tract model. 
The model is built from MRI data of a German male speaker, and is controlled by 20 parameters. 

\begin{figure}[]
  \centering
  \includegraphics[width=.49\linewidth]{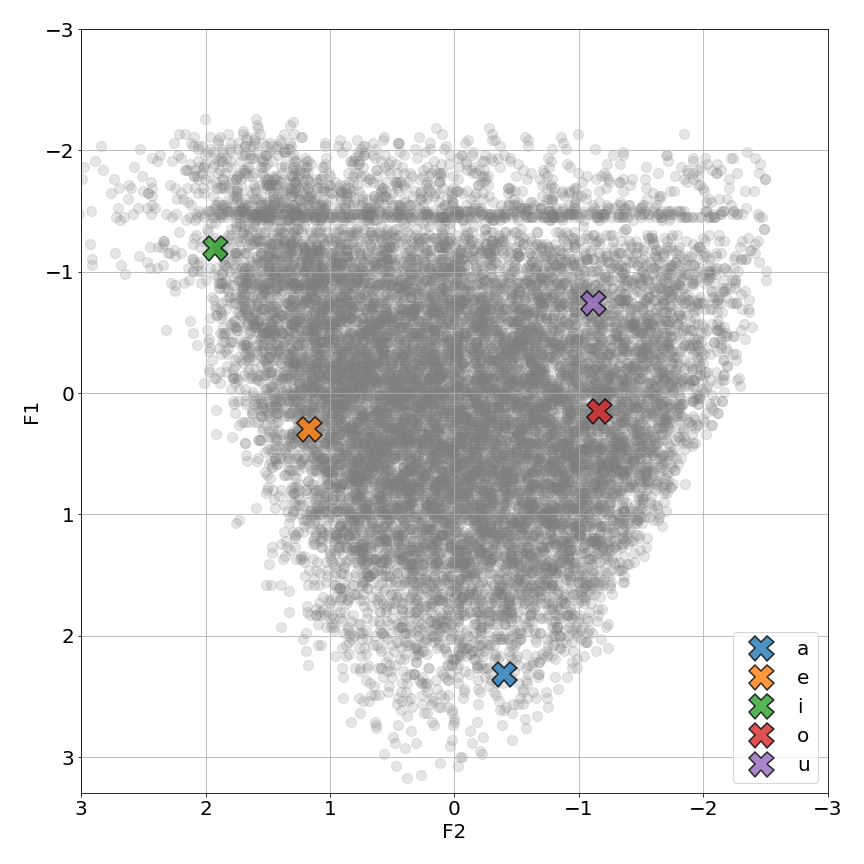}
  \includegraphics[width=.49\linewidth]{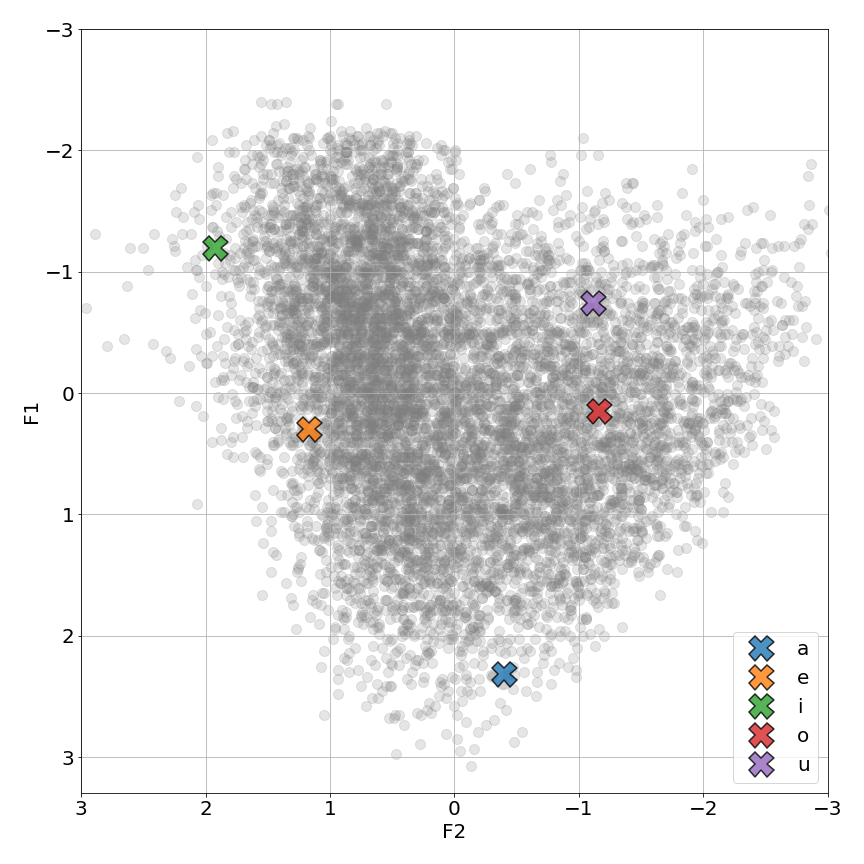}
  \caption{Formant spread of the synthesised VTSs for the adult (left) and the child model (right) in normalised F1-F2 space. Target vowels are represented with colour markers.}
  \label{fig:data_formants}
\vspace{-10pt}
\end{figure}

\noindent\textbf{Synthesised data.}
Two models were used in the analysis: 
\emph{i}) the original adult model based on the MRI scans of a human subject, and 
\emph{ii}) a prototype child model created as a scaled down version of the adult model~\cite{birkholz2007simulation}.

We generated in total 1 million vocal tract shapes (VTSs) for both models by random sampling of 17 of the 24 VTL vocal tract parameters in the parameter ranges of the speaker models:
hyoid $x$ and $y$ position (HX, HY), jaw $x$ and angle (JX, JA), lip protrusion and distance (LP, LD), velum shape (VS), tongue centre, blade and tip $x$ and $y$ (TCX, TCY, TBX, TBY, TTX, TTY), and the four tongue side vertical positions (TS1, TS2, TS3, TS4) \cite{birkholz2013modeling}.
We generated the 500,000 VTSs for each model in 5 runs with 100,000 iterations each.
All runs started from the neutral vocal tract position corresponding to a central schwa.

We prefiltered the VTSs based on the positional constraints for the tongue parameters
and vocal tract closure.
We then extracted F1 and F2 from the magnitude of the volume velocity transfer function using a peak picking algorithm
and
postfiltered the VTSs based on the expected F1 and F2 ranges
\cite{lee1997analysis}.
%
Finally, we postfiltered the synthesised speech signals based on their low-frequency energy to include only VTSs that allowed sustained phonation with VTL's acoustical coupling.

This rigorous selection process resulted with 15,229 (3\% of the original VTS samples) for the adult model and 8,510 (1.7\%) for the child model.\footnote{Supplementary materials -- \url{http://www.homepages.ucl.ac.uk/~uclyyix/EVL/feature-metric.html}}
Fig.~\ref{fig:data_formants} shows the formant spread for the two speaker models with the target vowel's formant frequencies superimposed.
We can see a well formed vowel triangle in both cases, with a larger spread for the child model in line with
the increased variability seen in children
\cite{lee1997analysis}.

\textbf{Target vowel templates.}
The human speaker static vowel target templates comprise a single renditions of the five vowels: /a/, /e/, /i/, /o/, and /u/, as used in standard Macedonian, spoken by a native male speaker.
This limited set
provides ample coverage of the formant space as can be seen in Fig.~\ref{fig:data_formants}.

\subsection{Features and metrics}

Two well established speech features were extracted using LibROSA\footnote{LibROSA v.0.7.1 \url{https://librosa.github.io/}} \cite{mcfee2015librosa} – the Log Mel Spectrogram and the MFCCs.
The Mel filter bank used to extract the features in both cases comprised 26 filters with a maximum frequency of 10 kHz.
From these, 12 and 22 MFCCs were extracted.
MFCC12 was meant to emulate the usual ASR setup \cite{young2006htk}, while the richer MFCC22 was taken at the upper limit beyond which speaker specific information is captured \cite{ryant2014highly}.
In addition, we included high frequency emphasis through preemphasis and cepstral liftering, as commonly used in ASR.
Finally, we also applied Cepstral Mean and Variance Normalisation to the MFCC based features using the means and variances of the features extracted from the synthesised sounds with the final set of VTSs.
For the target speaker we used the recordings of the vowel targets.
For each feature type we calculated the errors using four distance measures: the Mean Square Error (MSE), the Cosine distance, and the Manhattan and Chebyshev distances as extremes of the Minkowski distance.
All of this amounted to a total of 40 feature-metric pairs.

\subsection{Formant error}
The formant errors were calculated using the Euclidean distance in the normalised F1-F2 space in order to compensate for
the differences in the formant space between the models and the target speaker.
We normalise the models' and the target's formant values using $z$-score normalisation based on the speaker specific means and standard deviations.
Some 300 additional realisations of the five vowels were used for extracting the target's formant statistics.

In order to gain a better estimate of the feature-metric pair performance, we also split the selected VTSs into their original 5 runs that start from the neutral schwa position.
Each split keeps ample coverage of the F1-F2 space akin to the one shown in Fig.~\ref{fig:data_formants}.
This gives us 5 error minima for each of the 5 vowels, or 25 formant errors in total for each feature-metric pair.

\subsection{Formant space error surface projection}
For each feature-metric pair and each of the target vowels we also calculate the error surface in the normalised formant space.
We use these error surfaces to gain additional insight on the way the error calculated with the metric in the feature space relates to the formant space.
We first calculate the error for every synthesised sound with each feature-metric pair for every vowel.
For each parameter combination we scale the errors to 1 by dividing them by the maximum error.
Next, we bin and average the errors in the F1-F2 space with 30 bins for each formant in the normalised range $-3$ to $3$.
We then use these average errors to calculate any missing data using cubic interpolation.

\begin{figure}[]
  \centering
  \includegraphics[width=\linewidth]{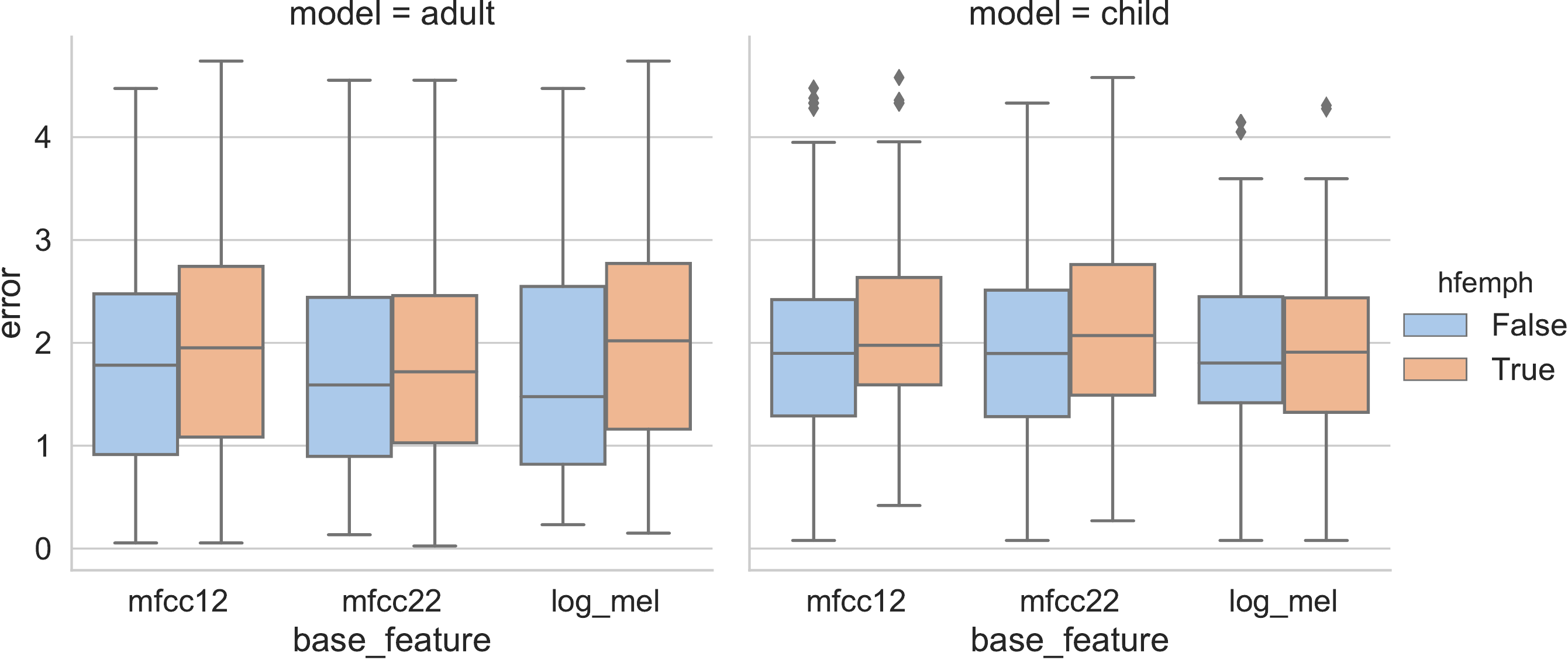}
  \caption{Aggregated impact of high-frequency emphasis.}
  \label{fig:hfemph}
\vspace{-10pt}
\end{figure}

\subsection{Listening tests}
To evaluate the perceptual relevance of the feature-metric pairs we design a MUSHRA (MUltiple Stimuli with Hidden Reference and Anchor) \cite{schoeffler2018webmushra} listening test in which we ask listeners to evaluate the phonetic accuracy of the synthesis that was selected as optimal by the feature-metric pairs for each vowel and each of the two models.
To optimise the listening tests we selected 10 of the feature-metric pairs based on their use in previous research and their formant error performance:
MFCC12 MSE, MFCC12 normalised MSE, MFCC12 COS, MFCC12 normalised COS, MFCC22 MSE, MFCC22 normalised MSE, MFCC22 COS, MFCC22 normalised COS, Log Mel spectrogram MSE, and Log Mel Chebyshev.
As negative anchors we use synthesised vowels different from the reference one.
We distributed the test to 10 speech researchers, of which 4 native speakers of Macedonian, and an additional 4 native speakers.
For each rater, we normalise the scores per model and vowel in the range 0 -- 1, using the scores given for the anchor and the reference.
We clip all negative scores to 0.

\begin{figure*}[t]
  \vspace{-5pt}
  \centering
  \includegraphics[width=.15\linewidth]{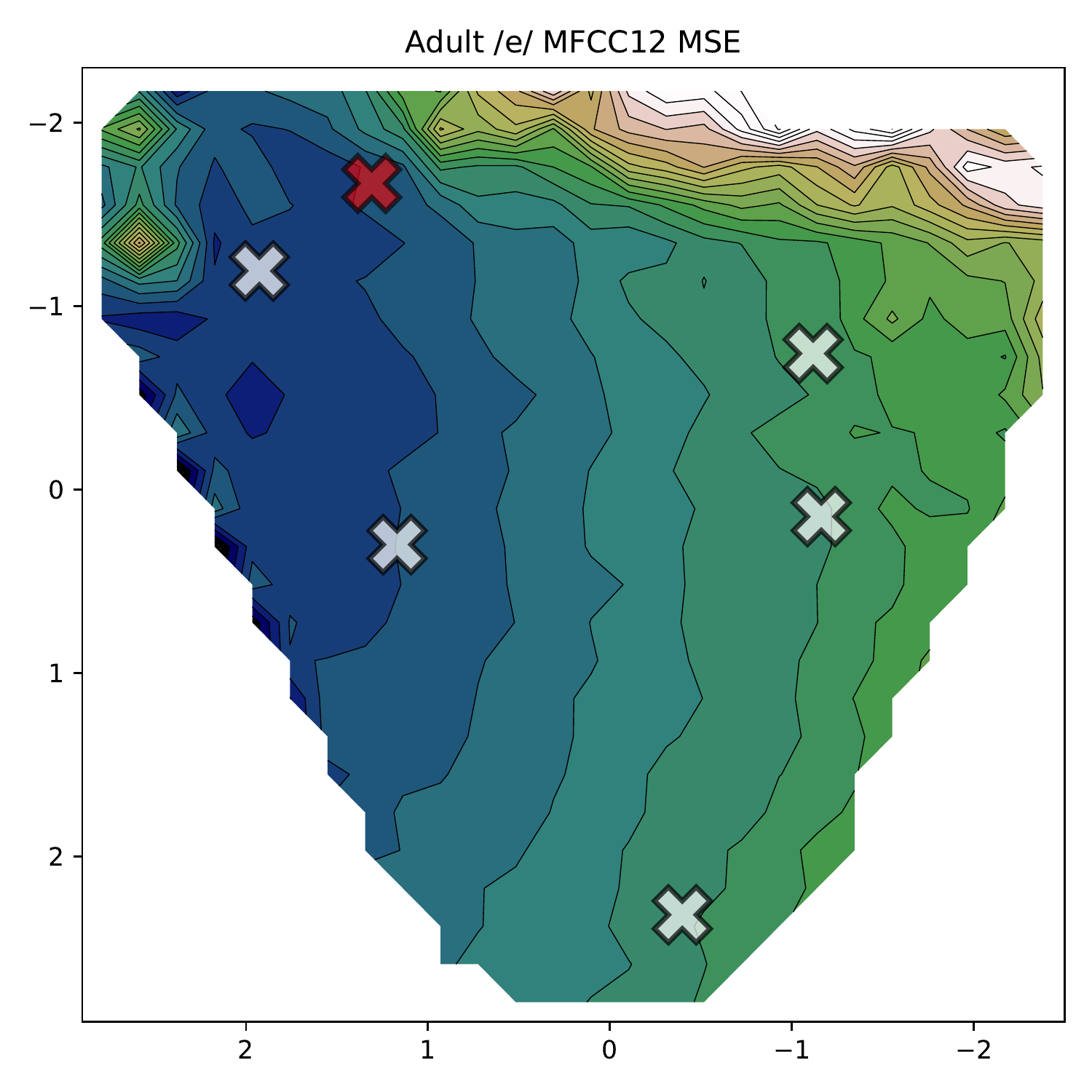}
  \includegraphics[width=.15\linewidth]{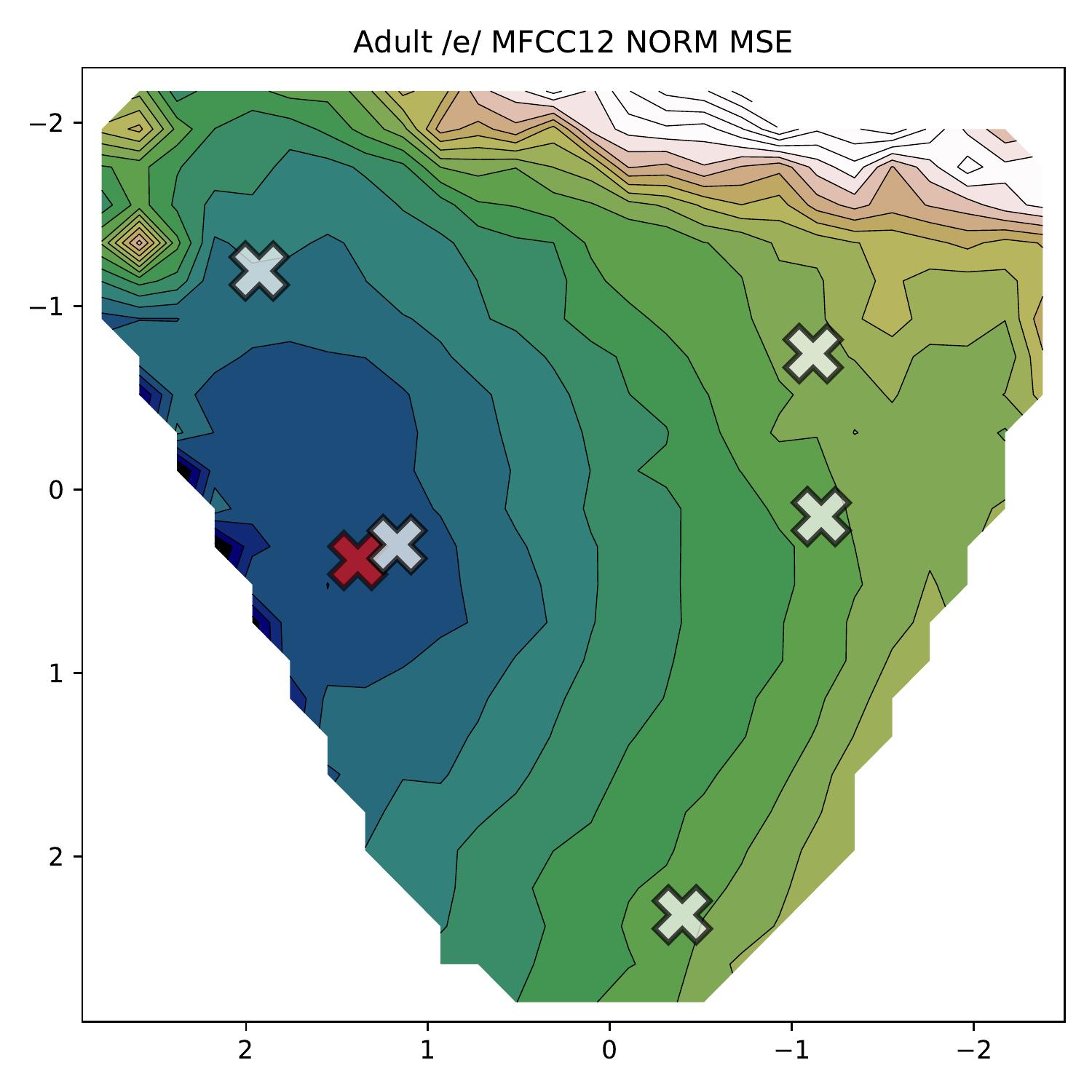}
  \includegraphics[width=.15\linewidth]{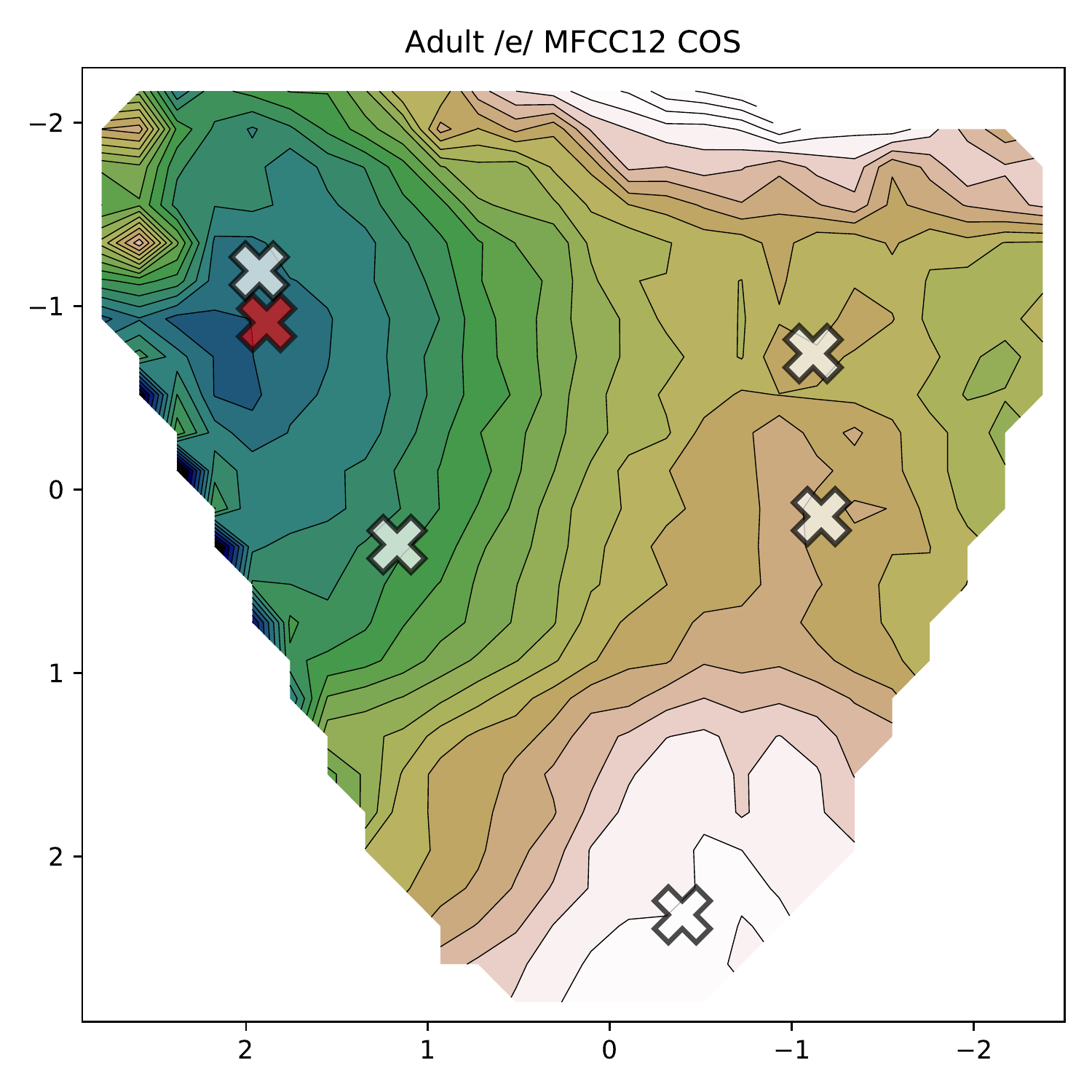}
  \includegraphics[width=.15\linewidth]{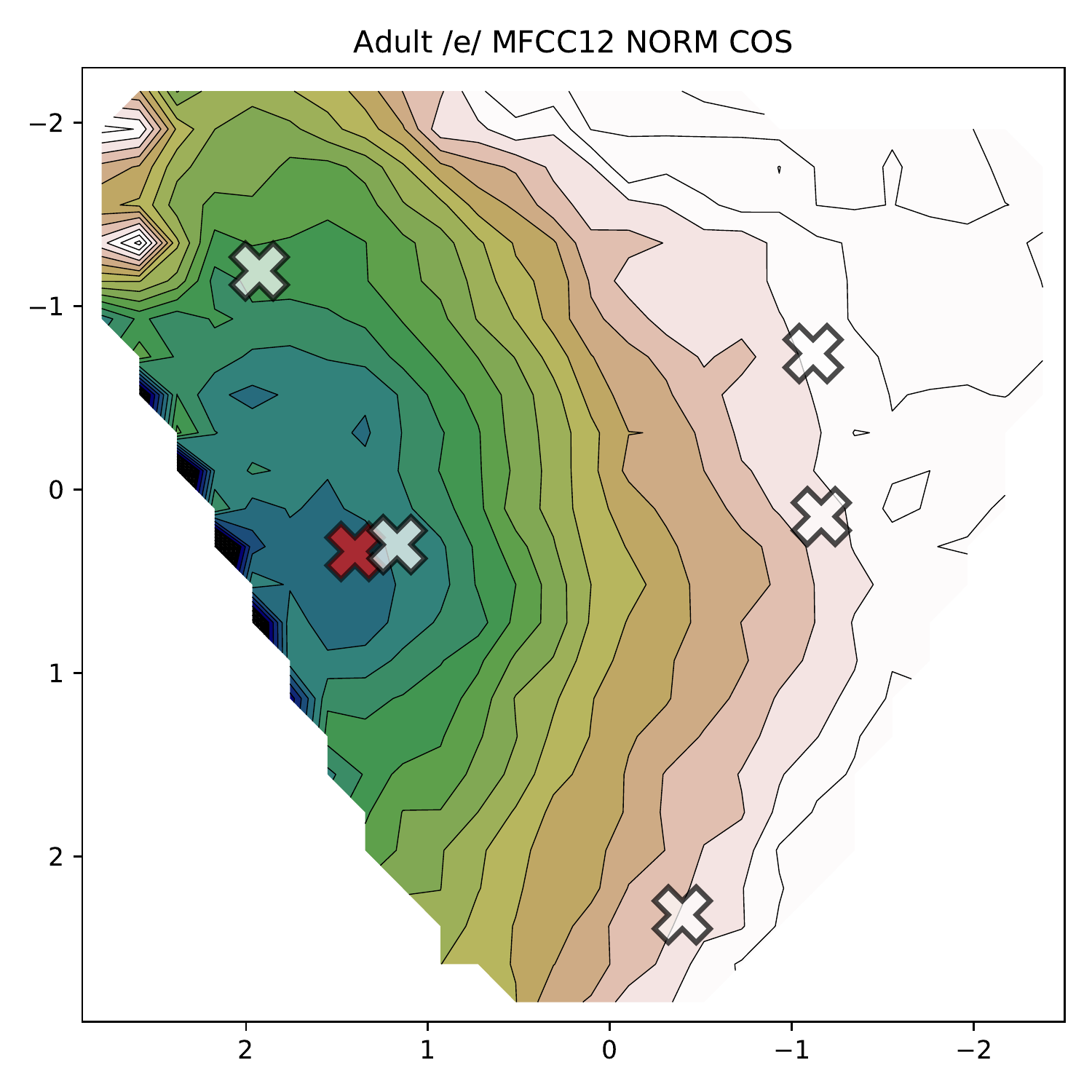}
  \includegraphics[width=.15\linewidth]{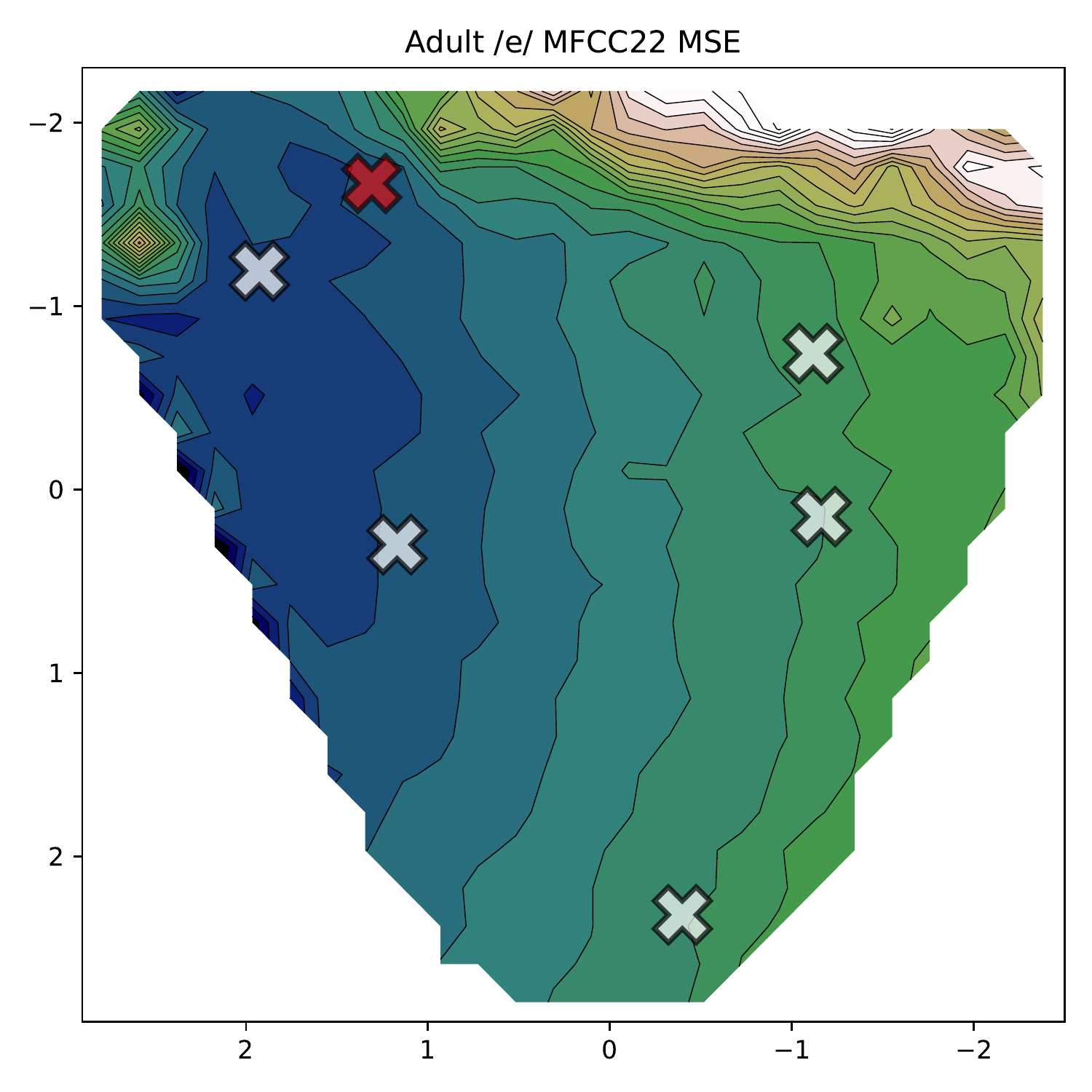}
  \includegraphics[width=.15\linewidth]{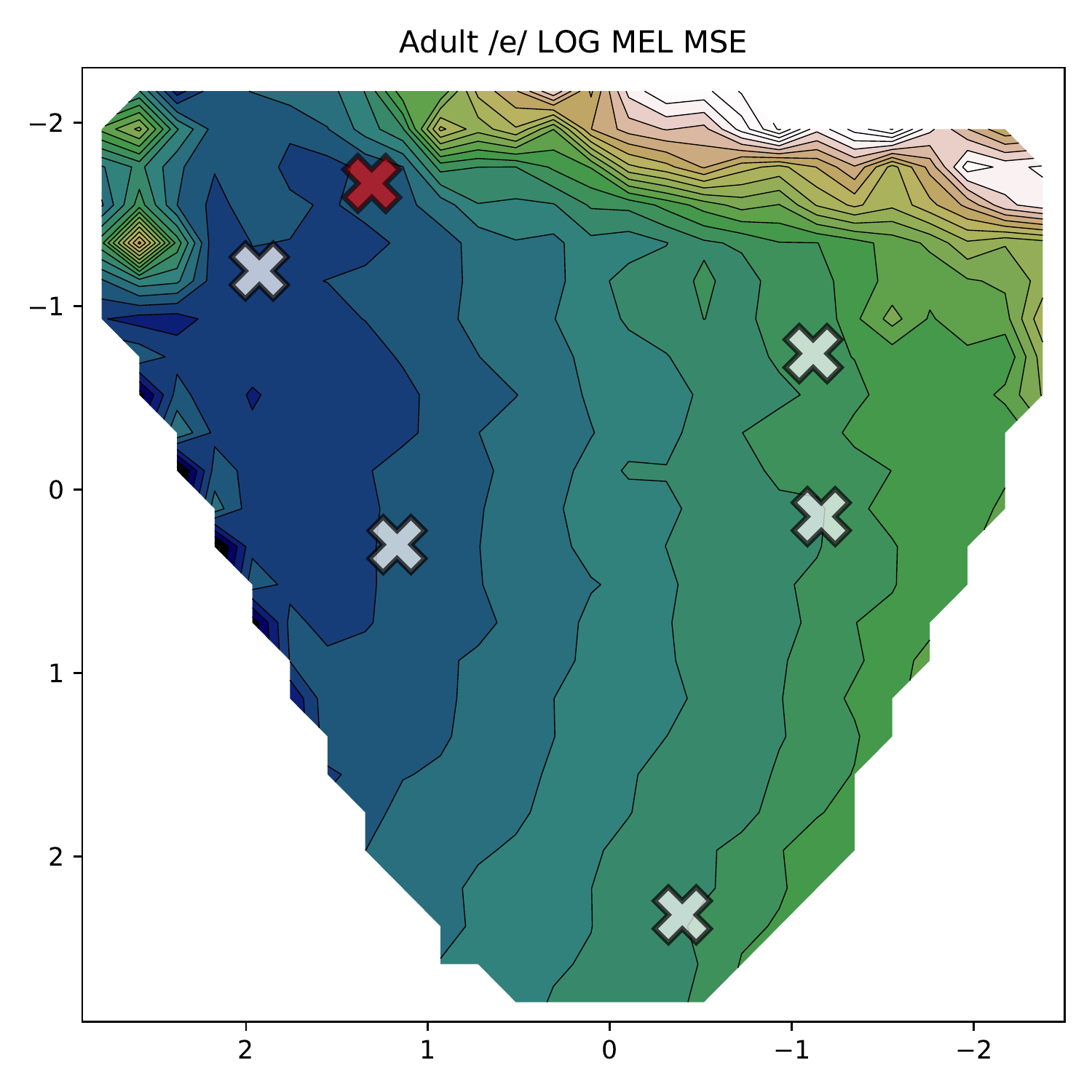}
  \includegraphics[width=.02\linewidth]{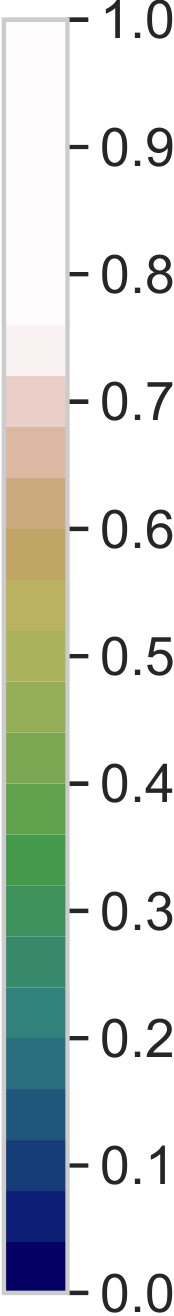}

  \includegraphics[width=.15\linewidth]{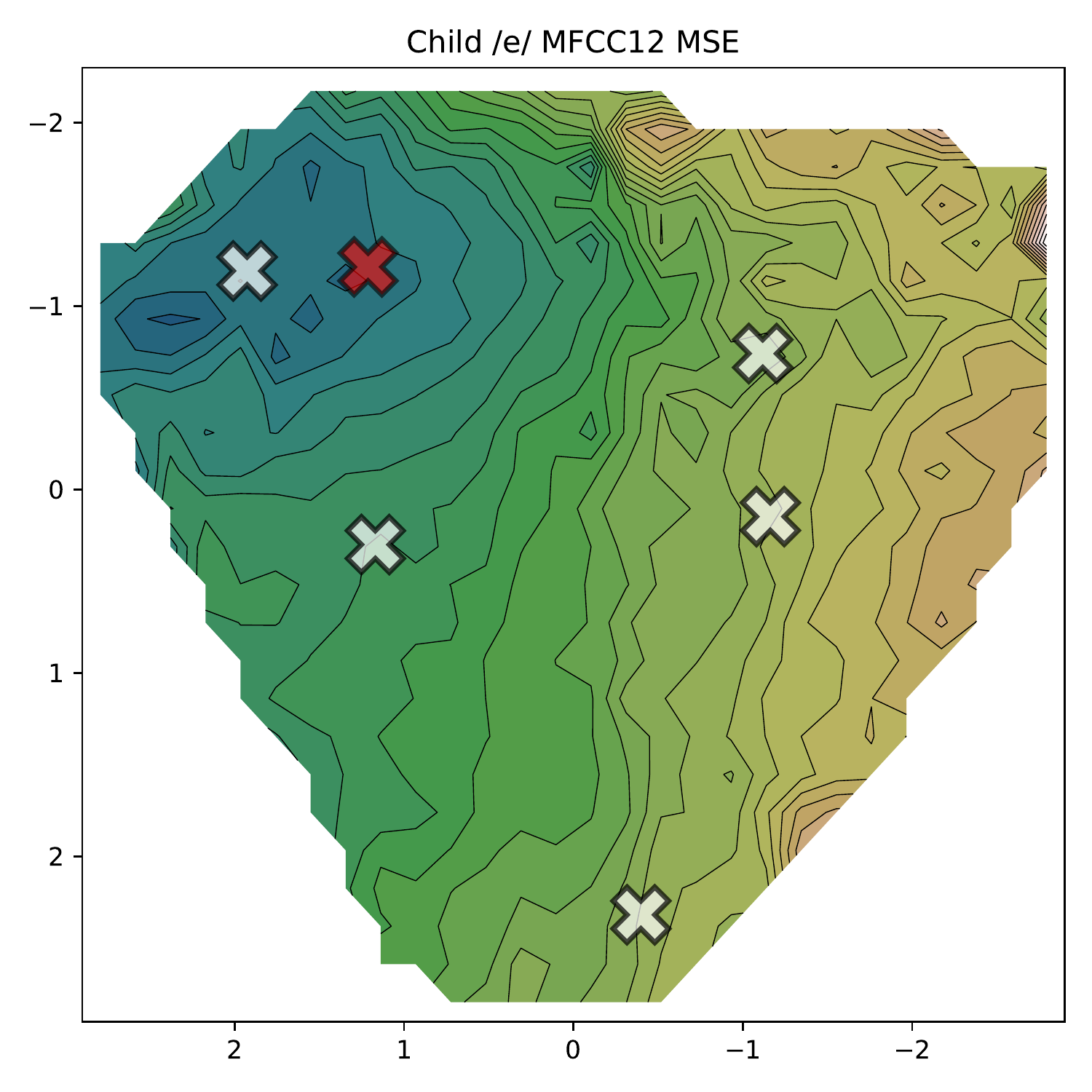}
  \includegraphics[width=.15\linewidth]{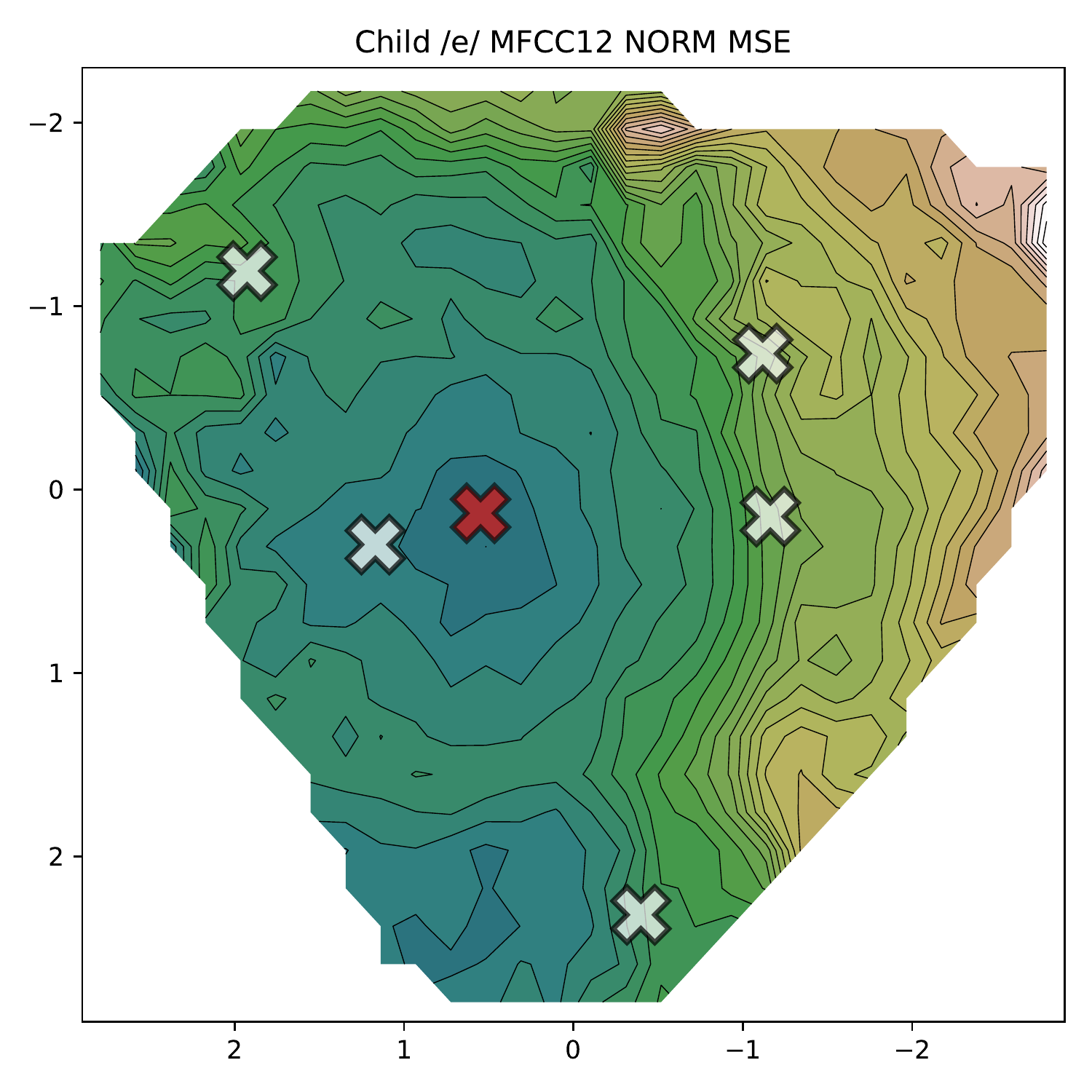}
  \includegraphics[width=.15\linewidth]{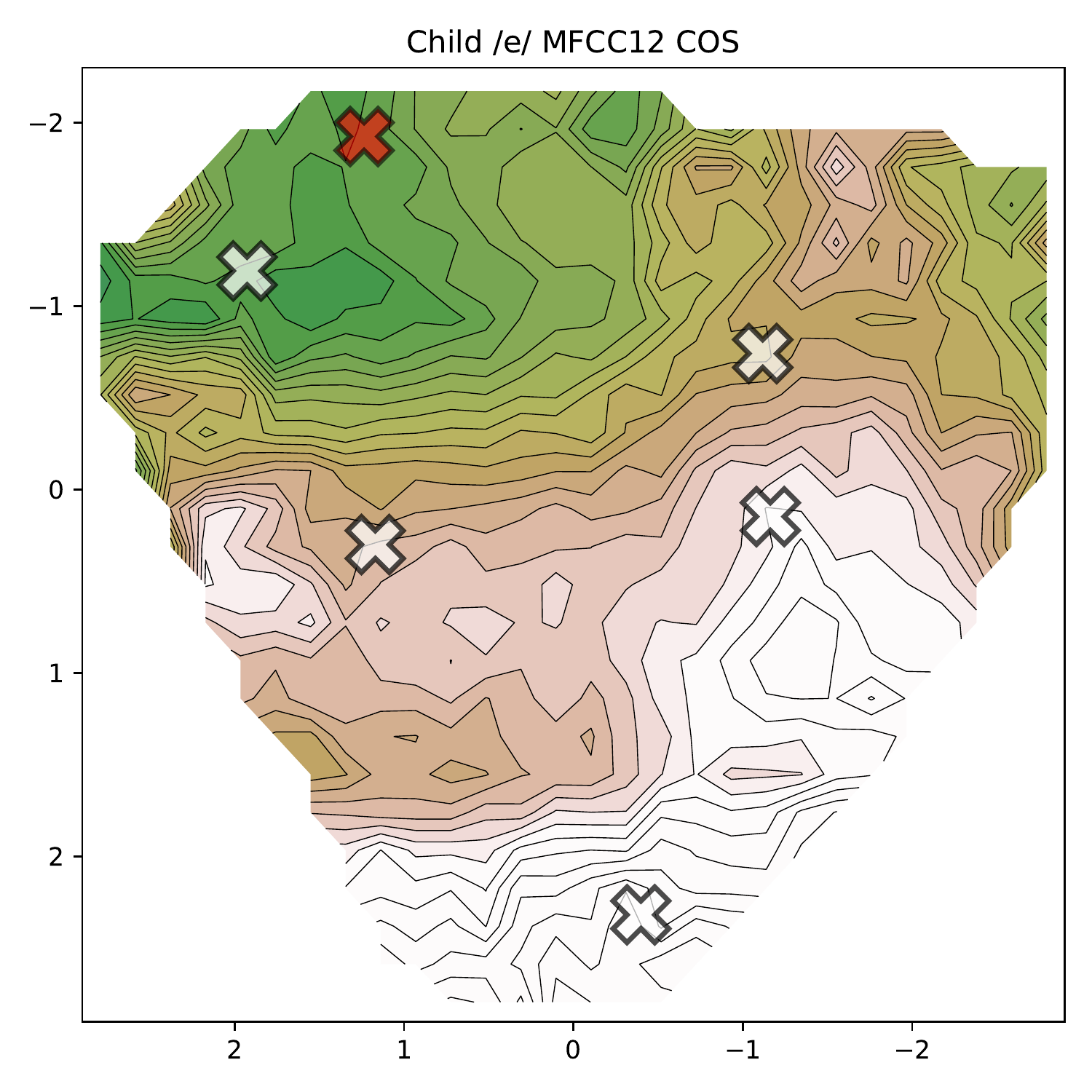}
  \includegraphics[width=.15\linewidth]{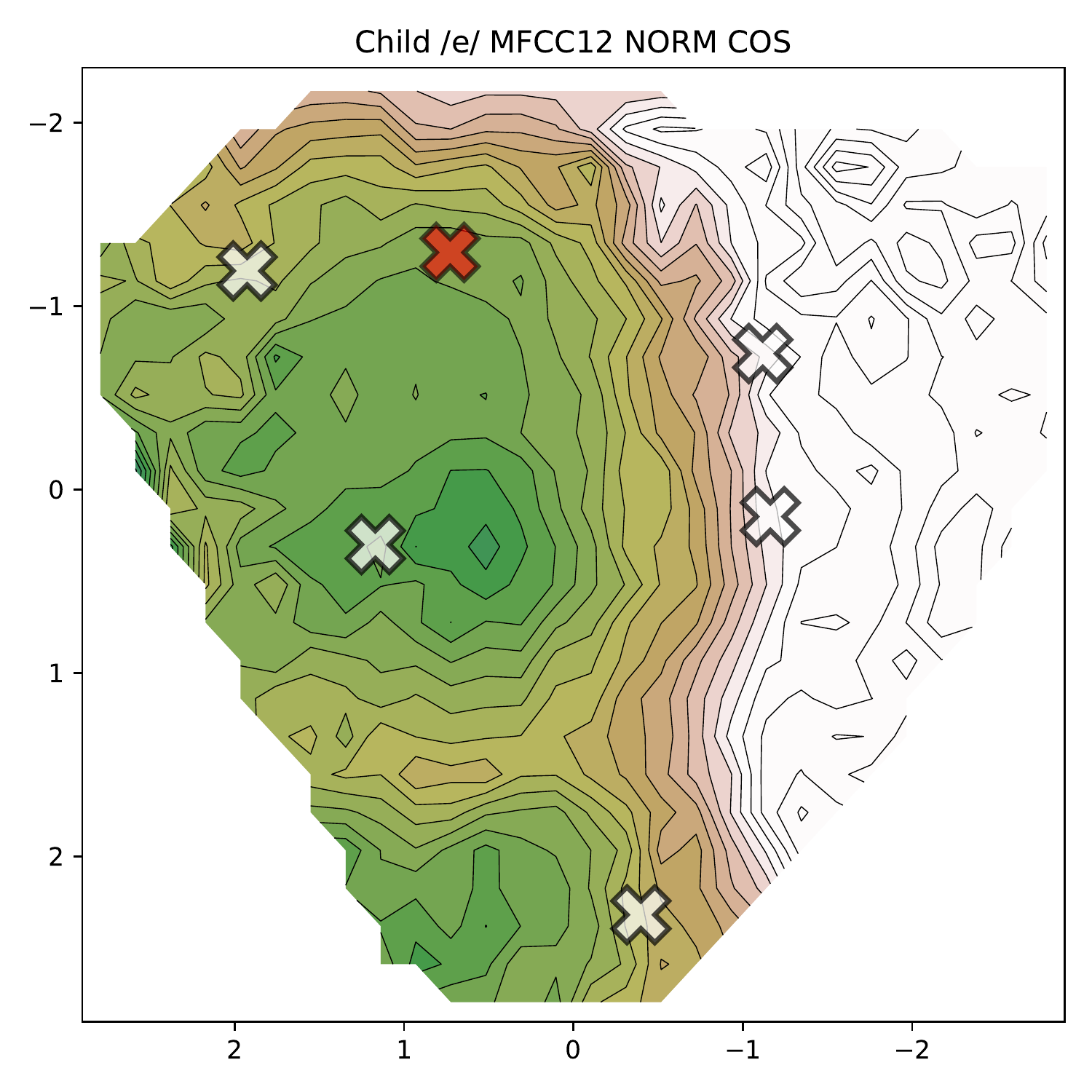}
  \includegraphics[width=.15\linewidth]{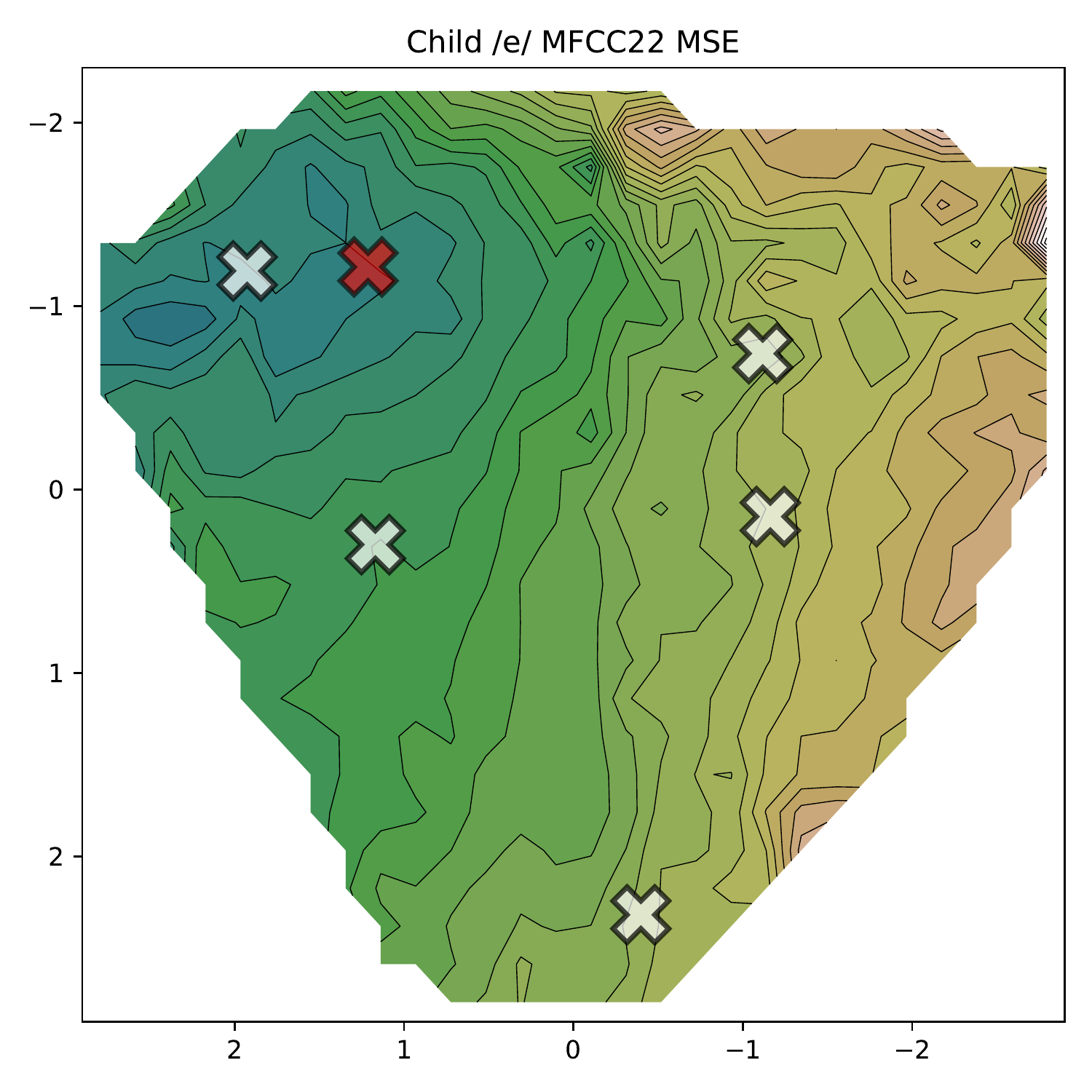}
  \includegraphics[width=.15\linewidth]{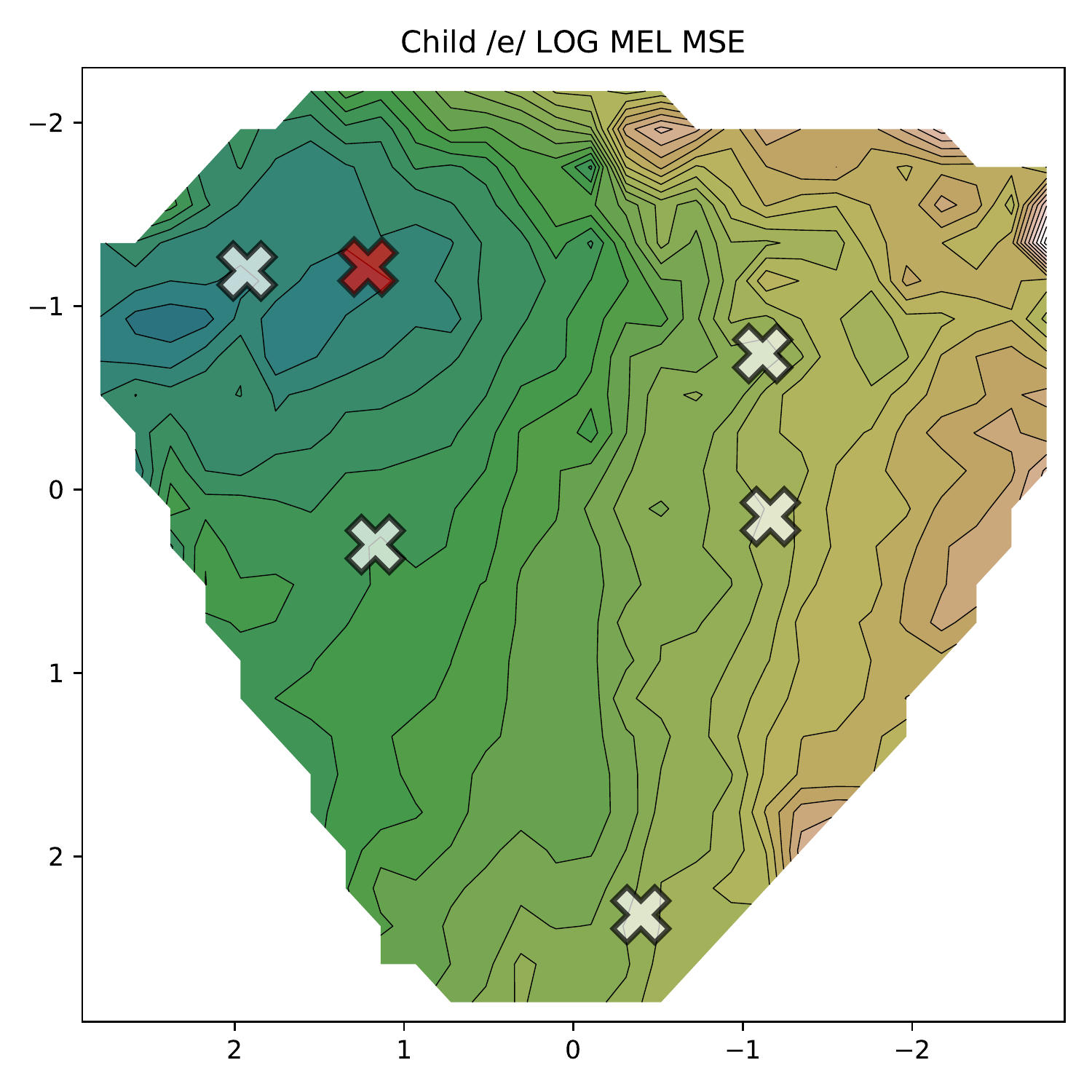}
  \hspace{.02\linewidth}

  \vspace{-5pt}
  \hrulefill
  \vspace{5pt}

  \includegraphics[width=.15\linewidth]{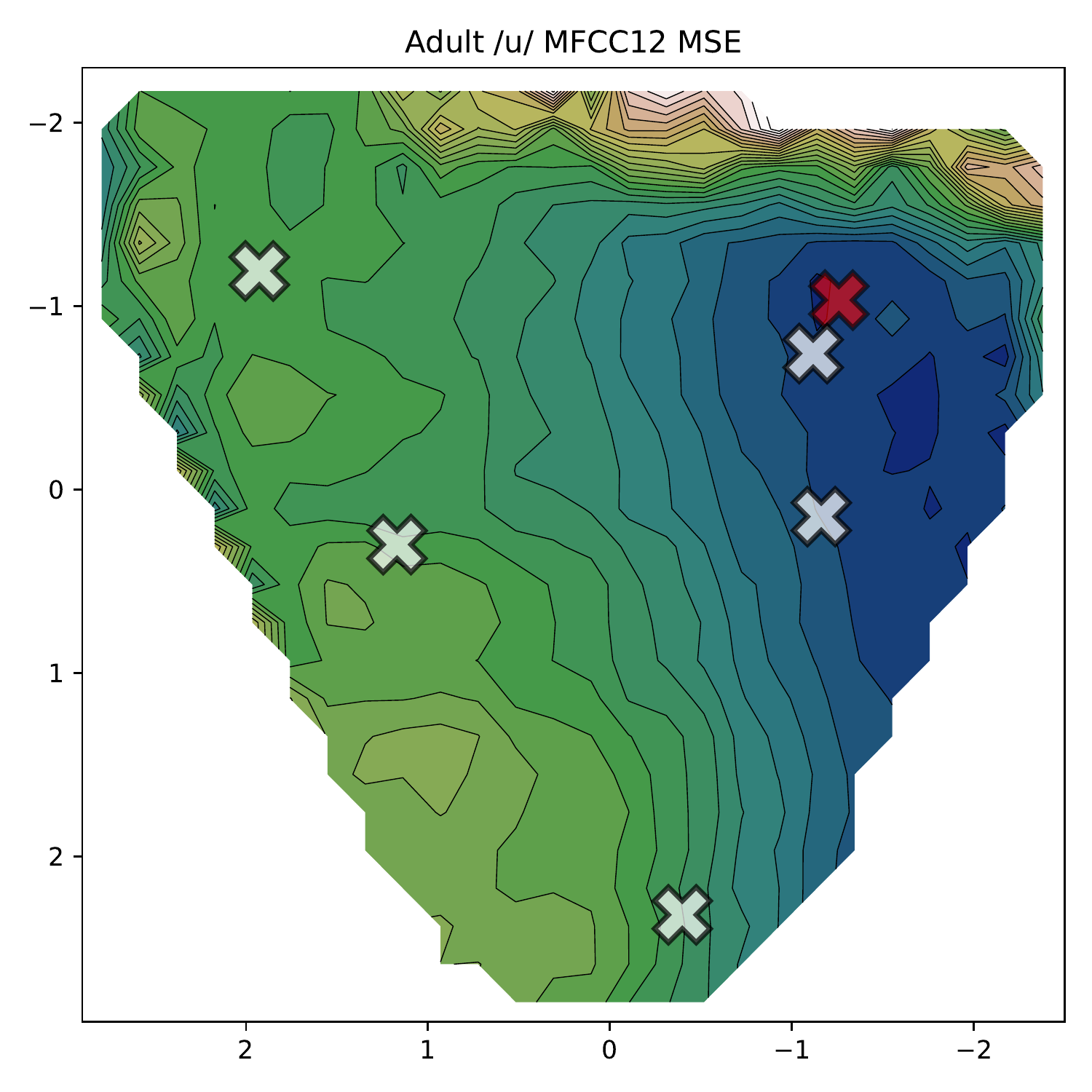}
  \includegraphics[width=.15\linewidth]{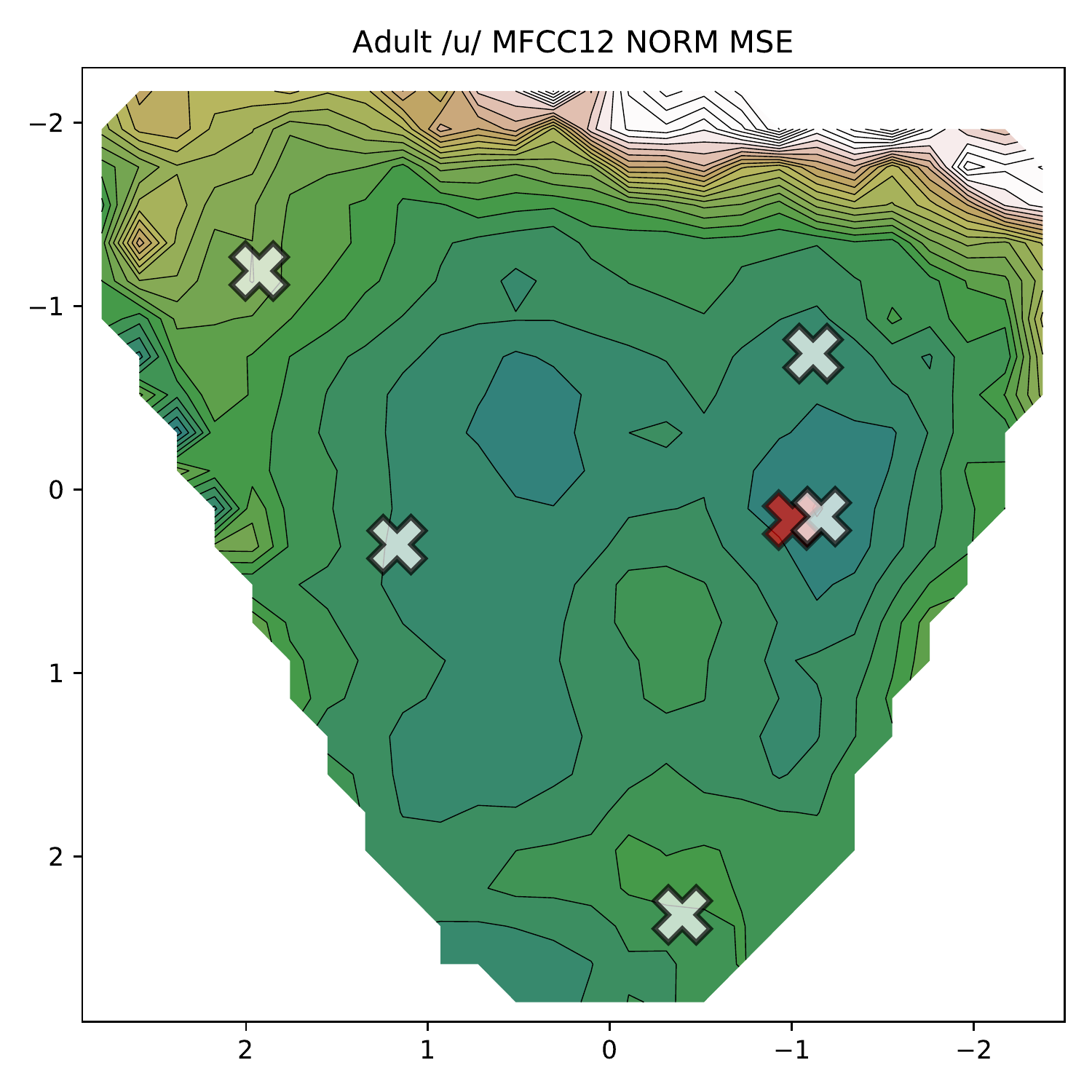}
  \includegraphics[width=.15\linewidth]{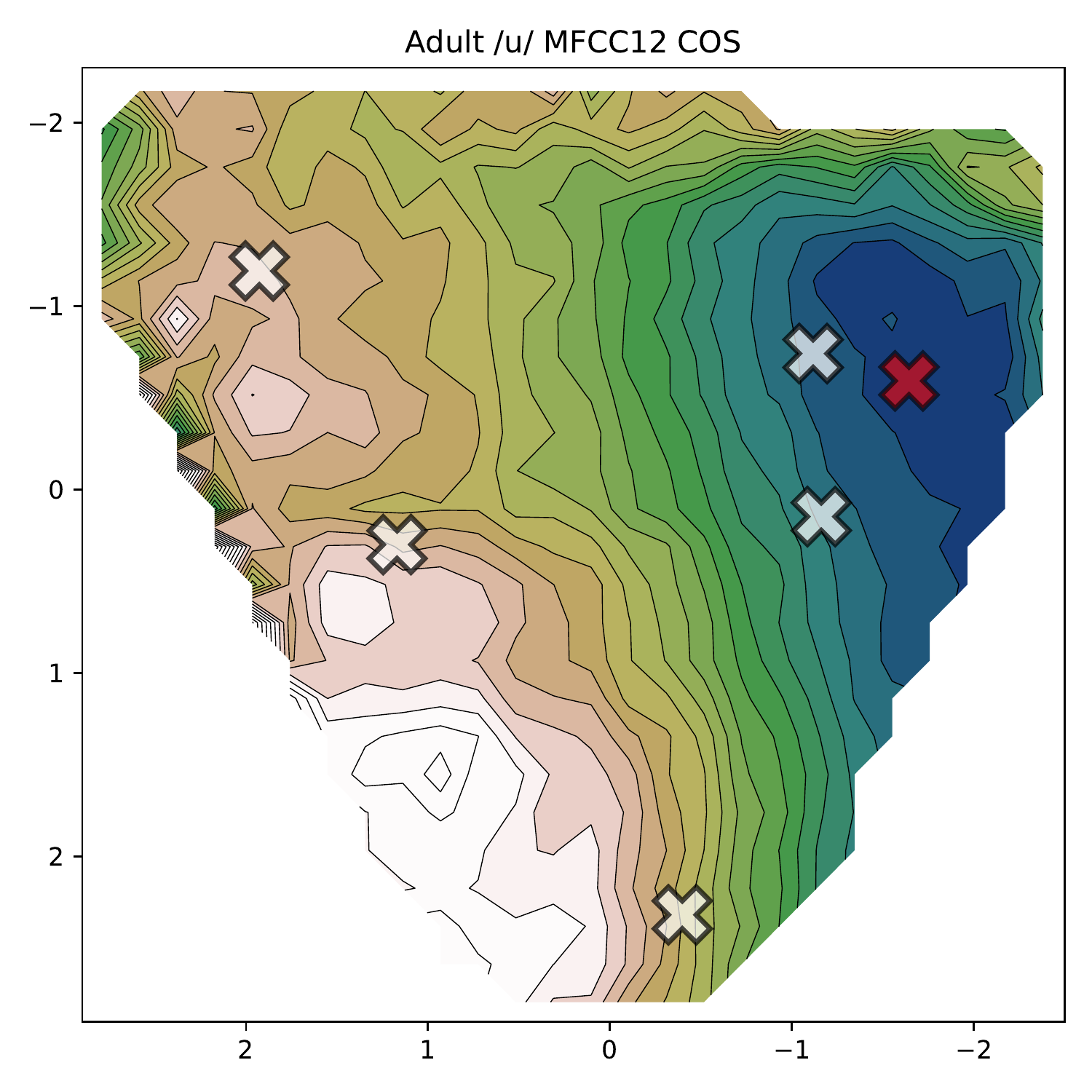}
  \includegraphics[width=.15\linewidth]{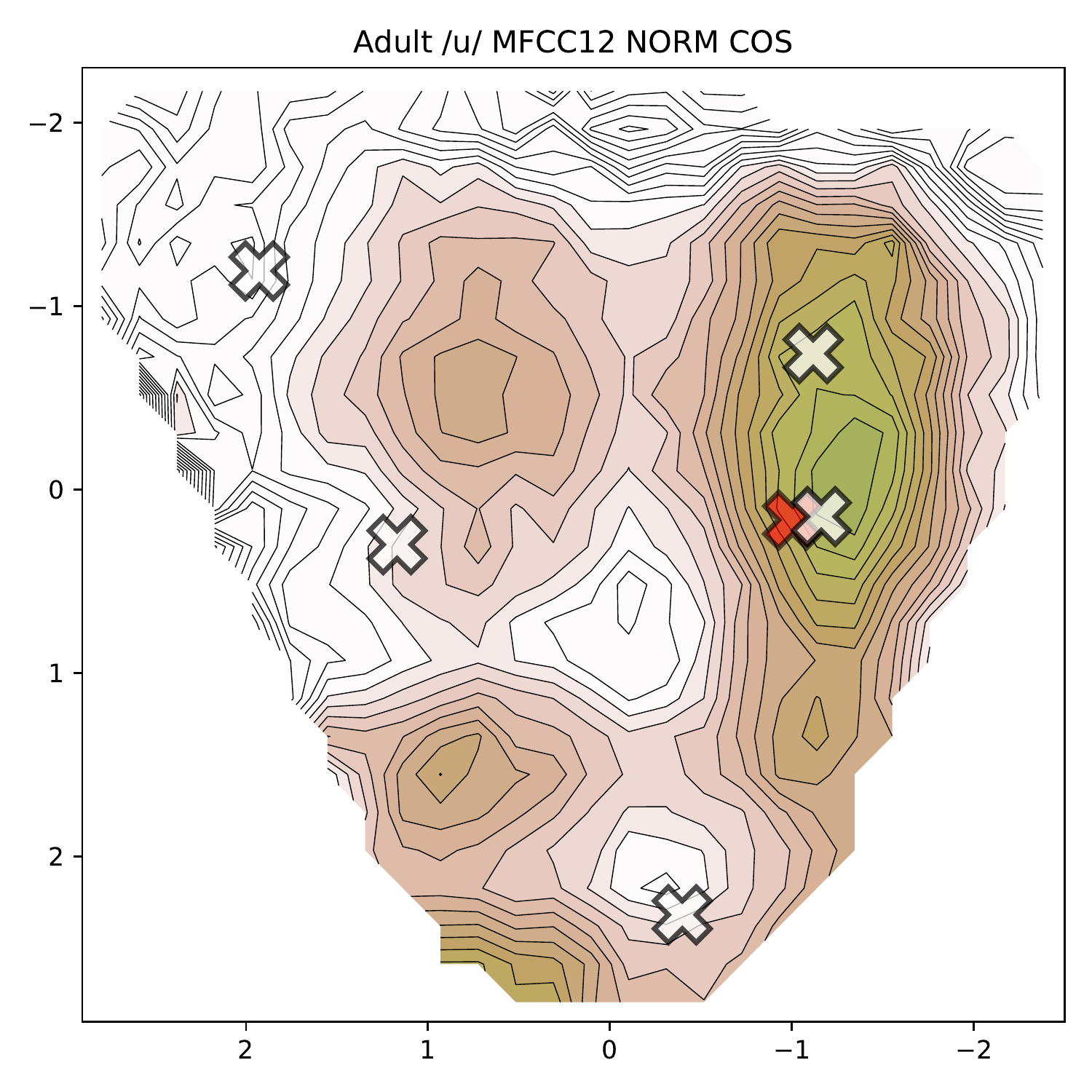}
  \includegraphics[width=.15\linewidth]{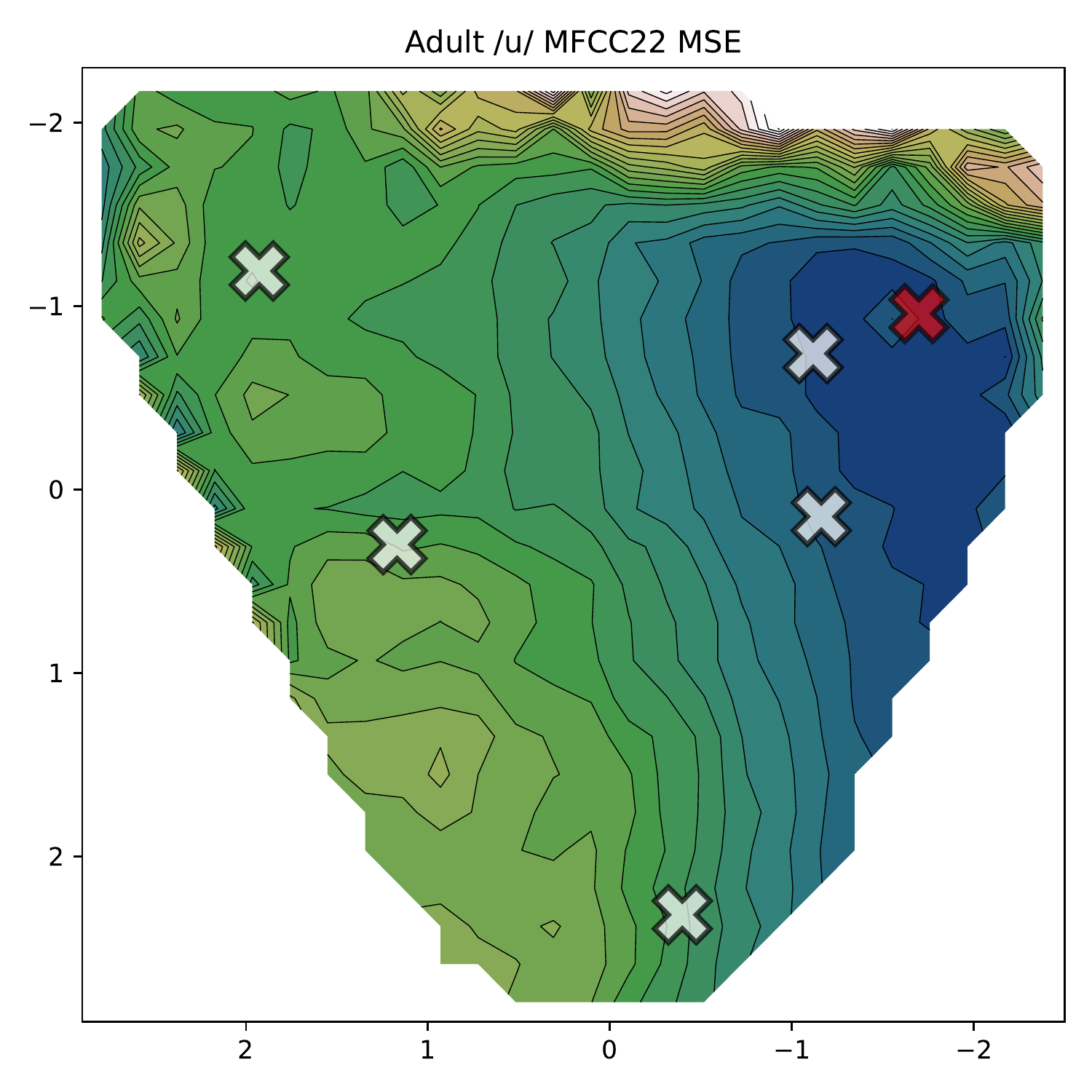}
  \includegraphics[width=.15\linewidth]{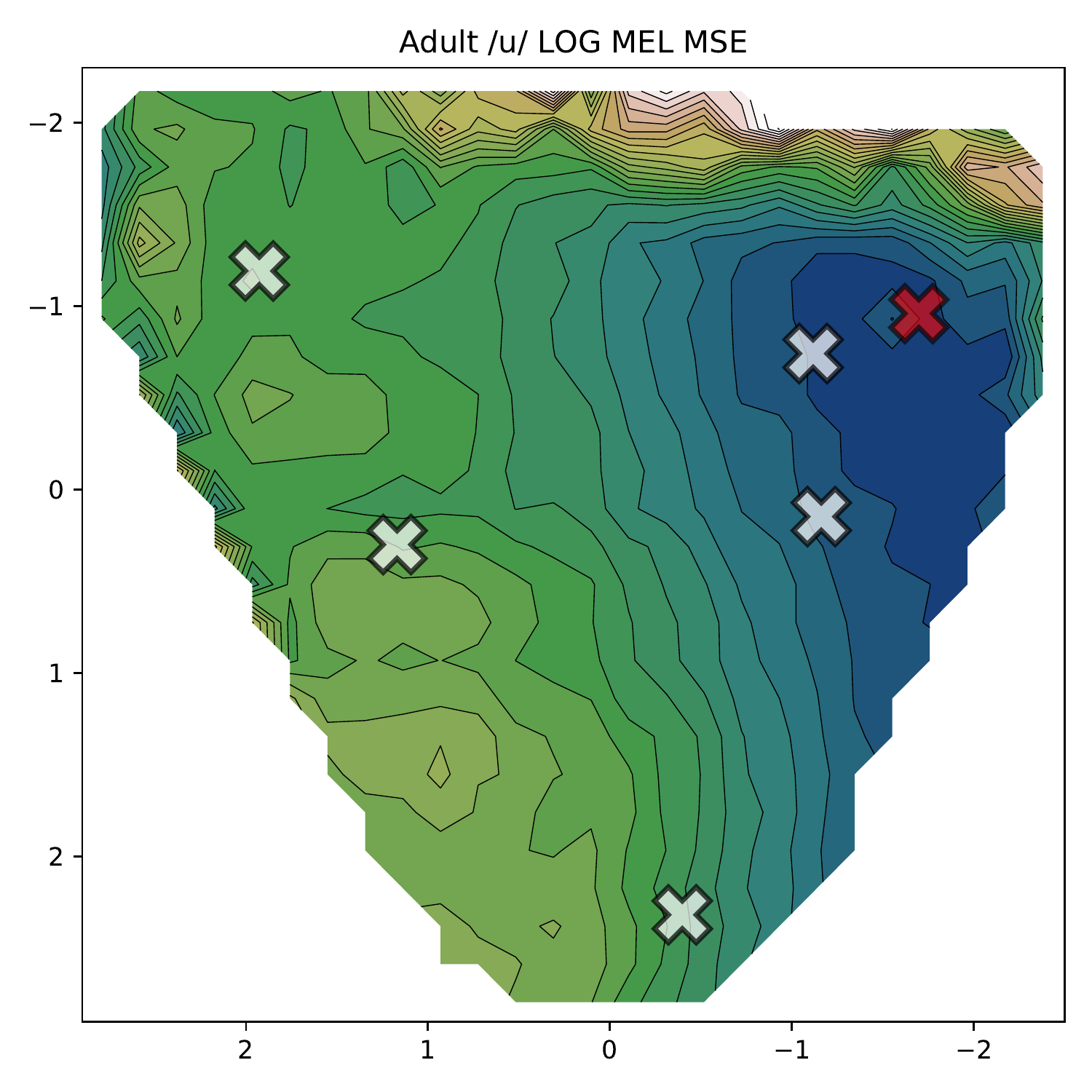}
  \includegraphics[width=.02\linewidth]{surf/colorbar.pdf}

  \includegraphics[width=.15\linewidth]{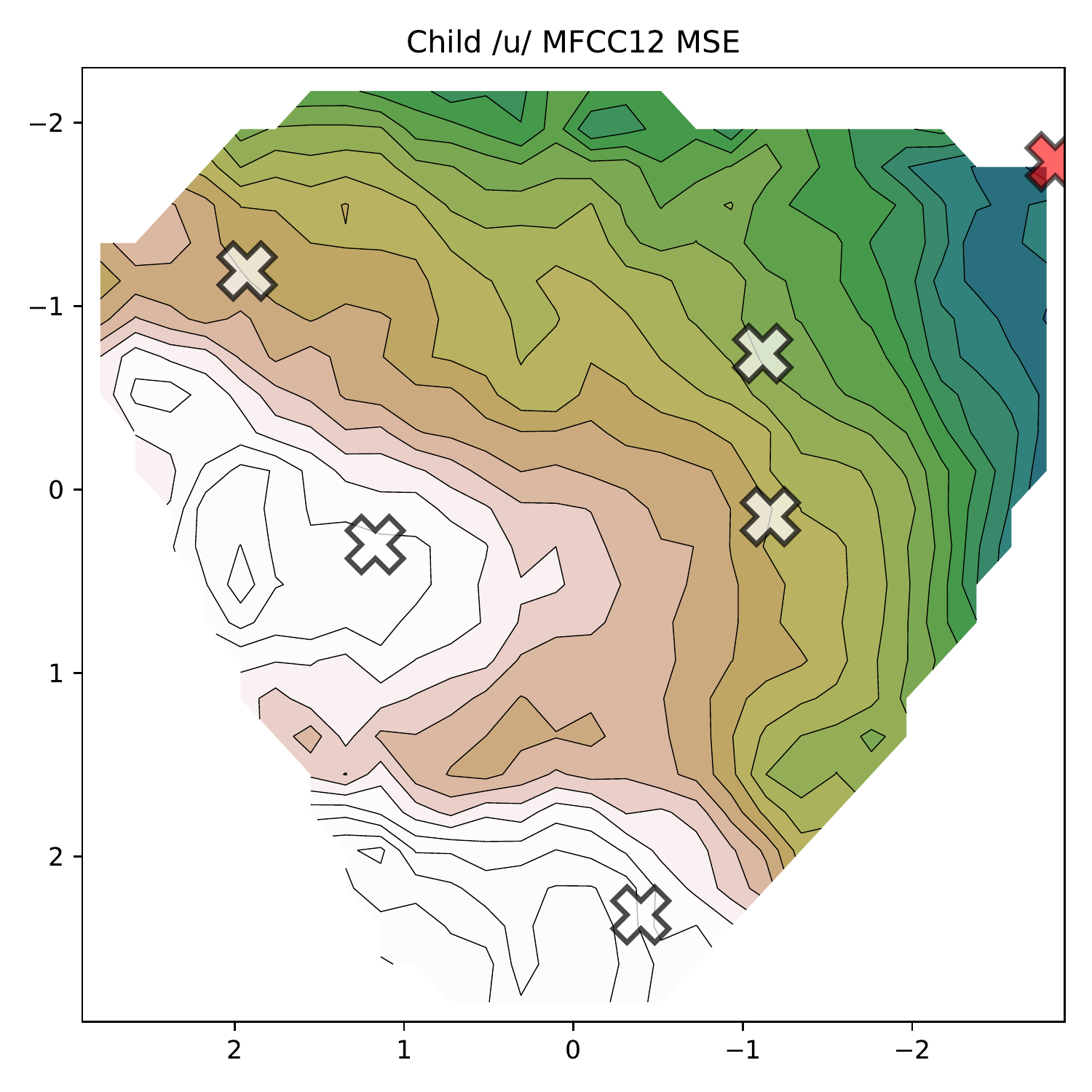}
  \includegraphics[width=.15\linewidth]{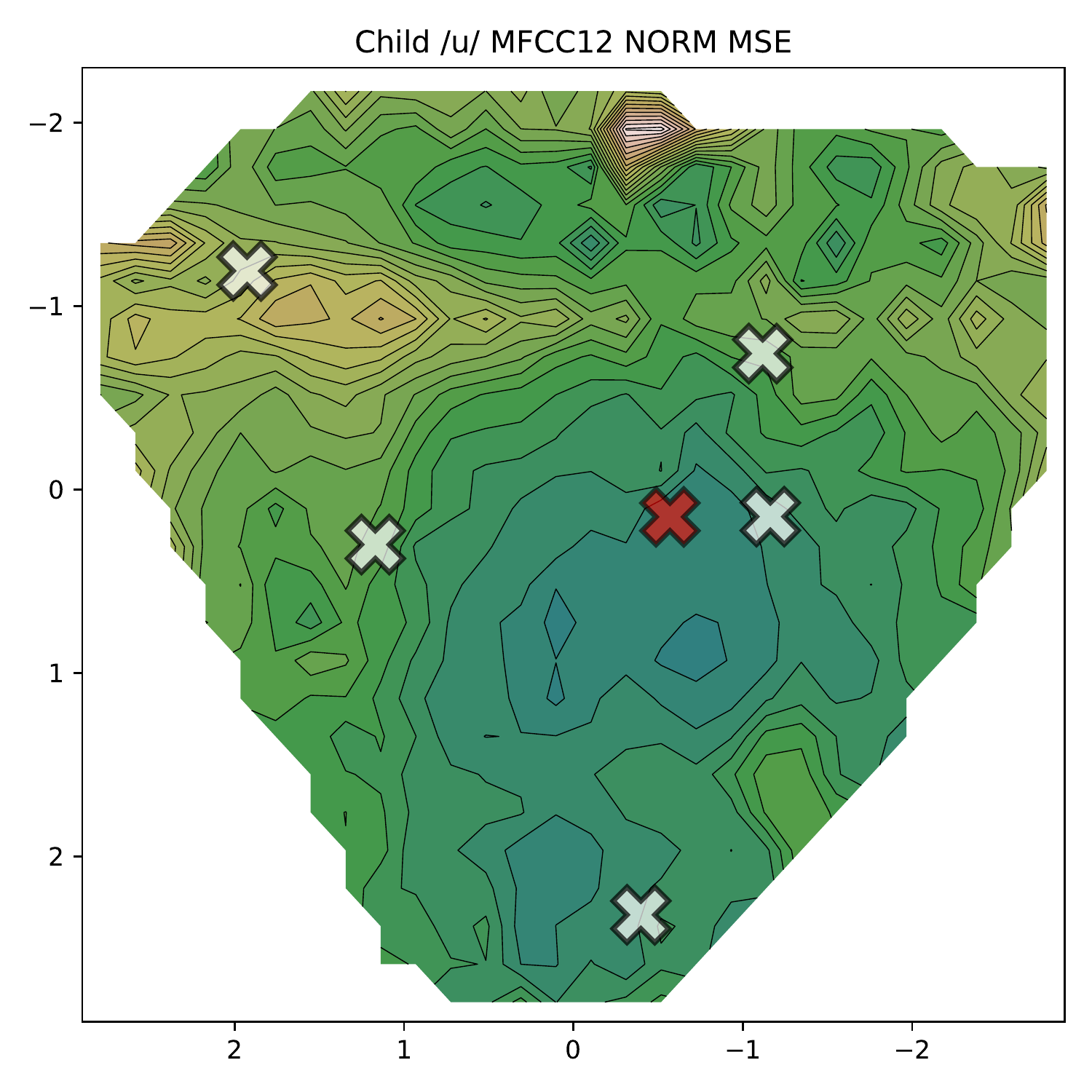}
  \includegraphics[width=.15\linewidth]{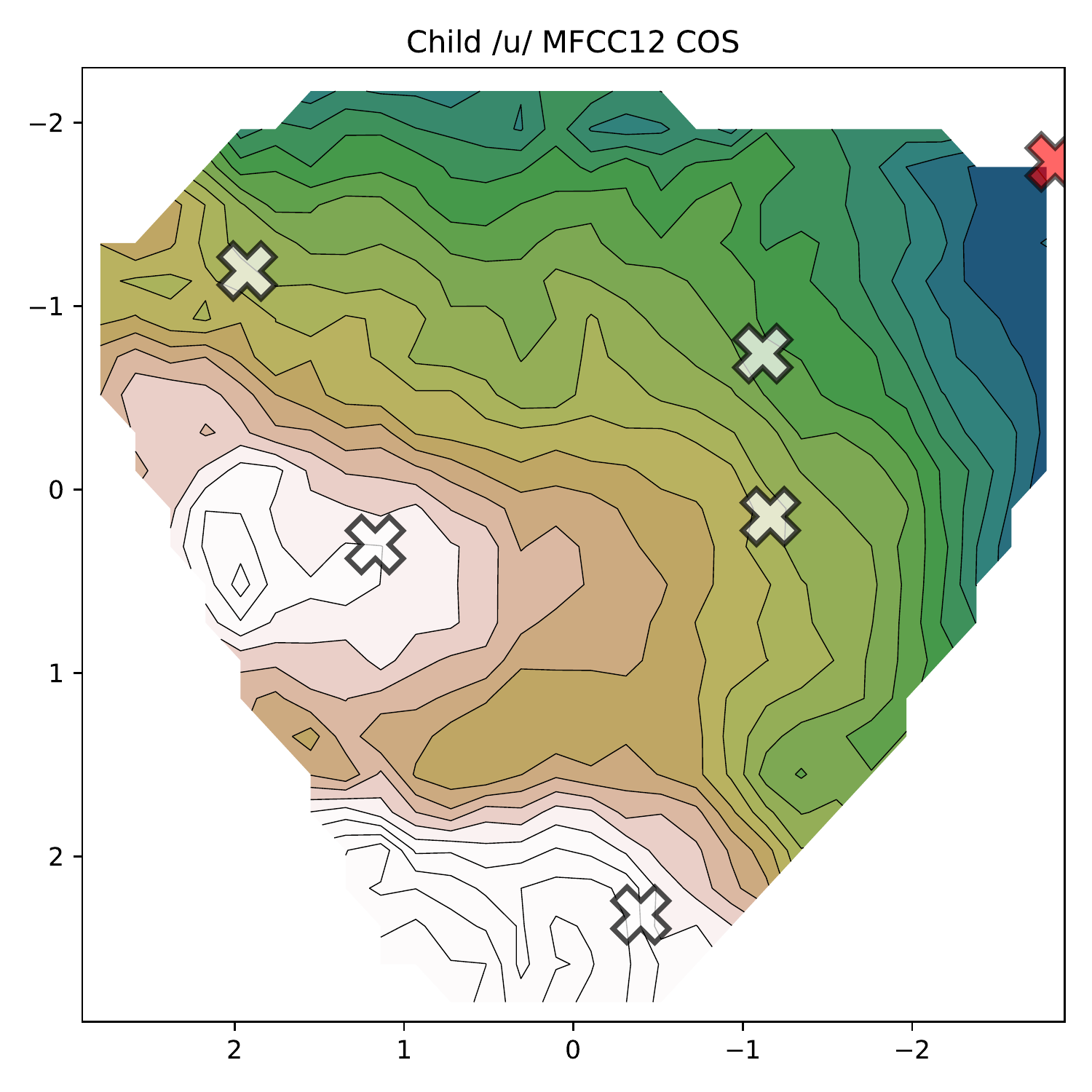}
  \includegraphics[width=.15\linewidth]{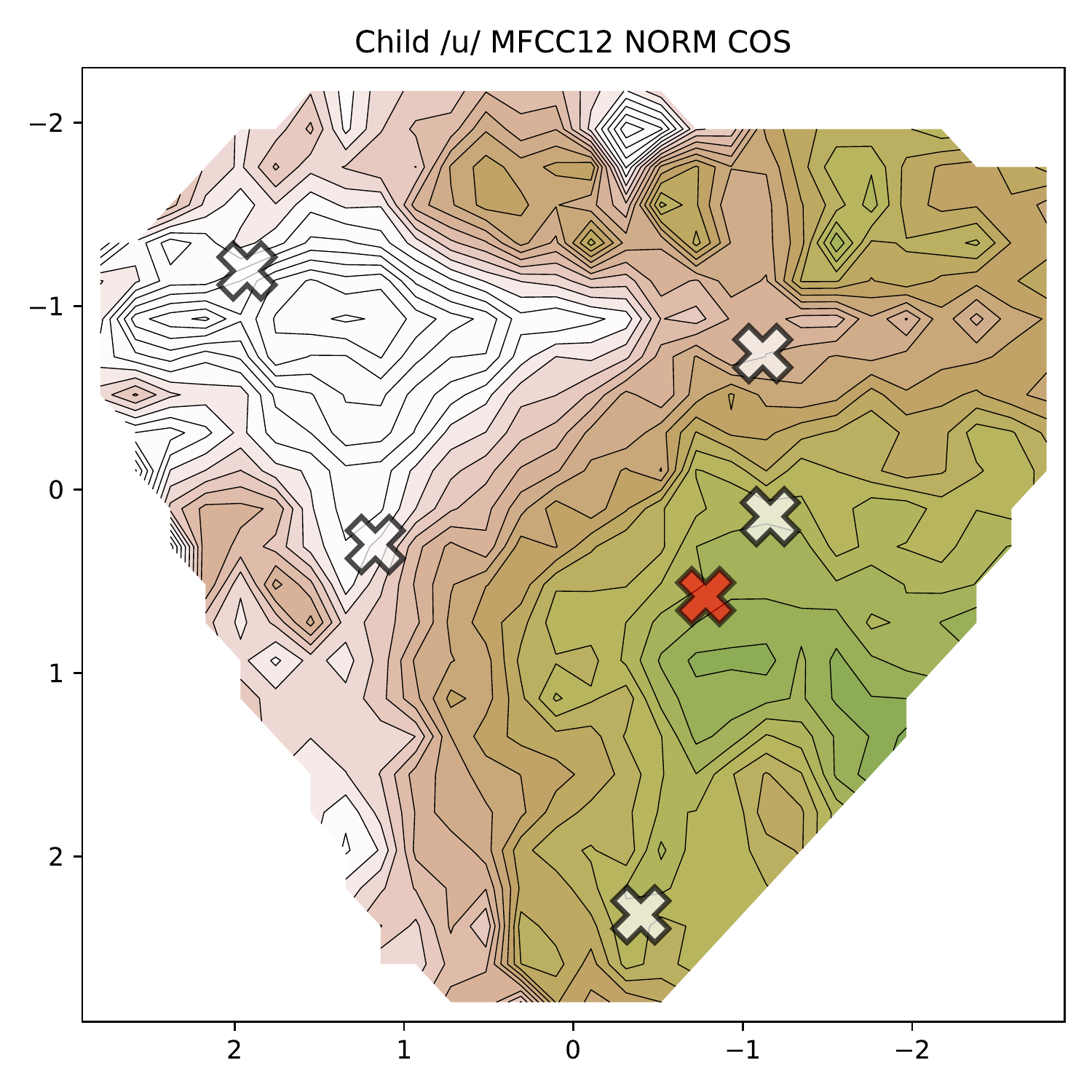}
  \includegraphics[width=.15\linewidth]{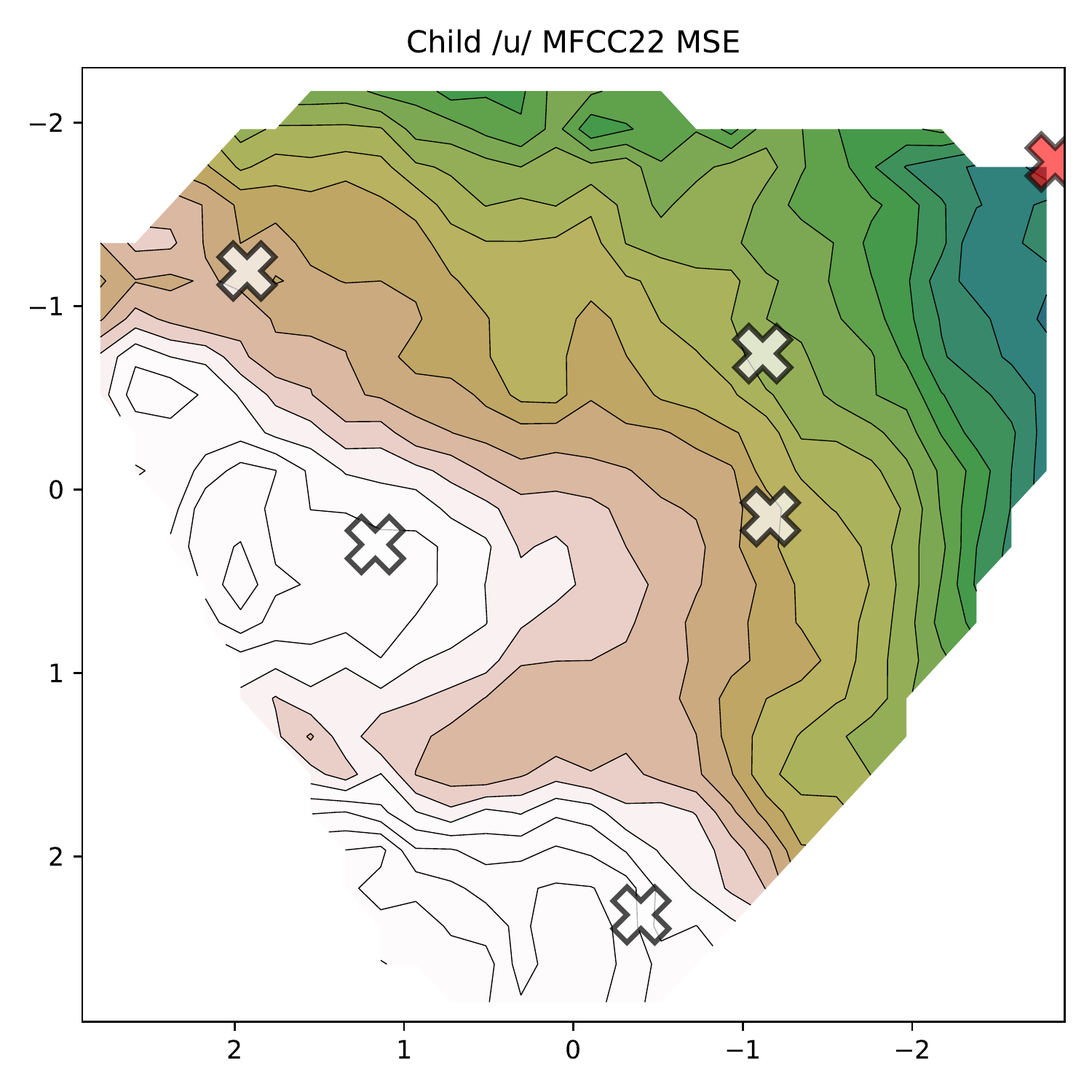}
  \includegraphics[width=.15\linewidth]{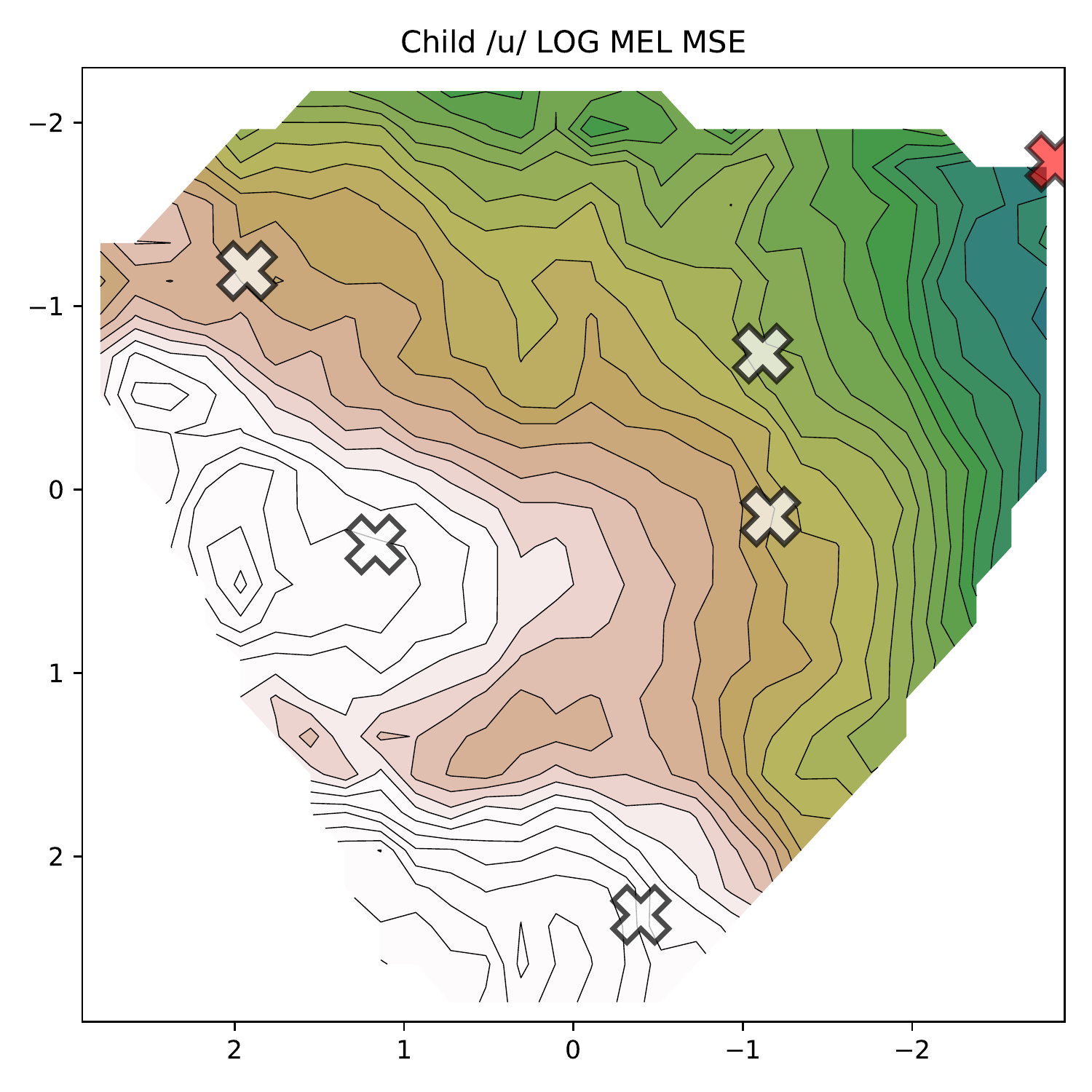}
  \hspace{.02\linewidth}

  \caption{MSE surface comparison for the target vowels /e/ (above line) and /u/ (below line) for the adult (top row) and child (bottom row) models for (left to right): MFCC12, MFCC12 N, MFCC12 COS, MFCC12 N COS, MFCC22 MSE, and Log Mel MSE. Target formants are superimposed with white markers and the formants of the signal with minimum error with a red marker.}
  \label{fig:surf_u}
  \vspace{-5pt}
\end{figure*}

\begin{figure}[]
  \centering
  \includegraphics[width=\linewidth]{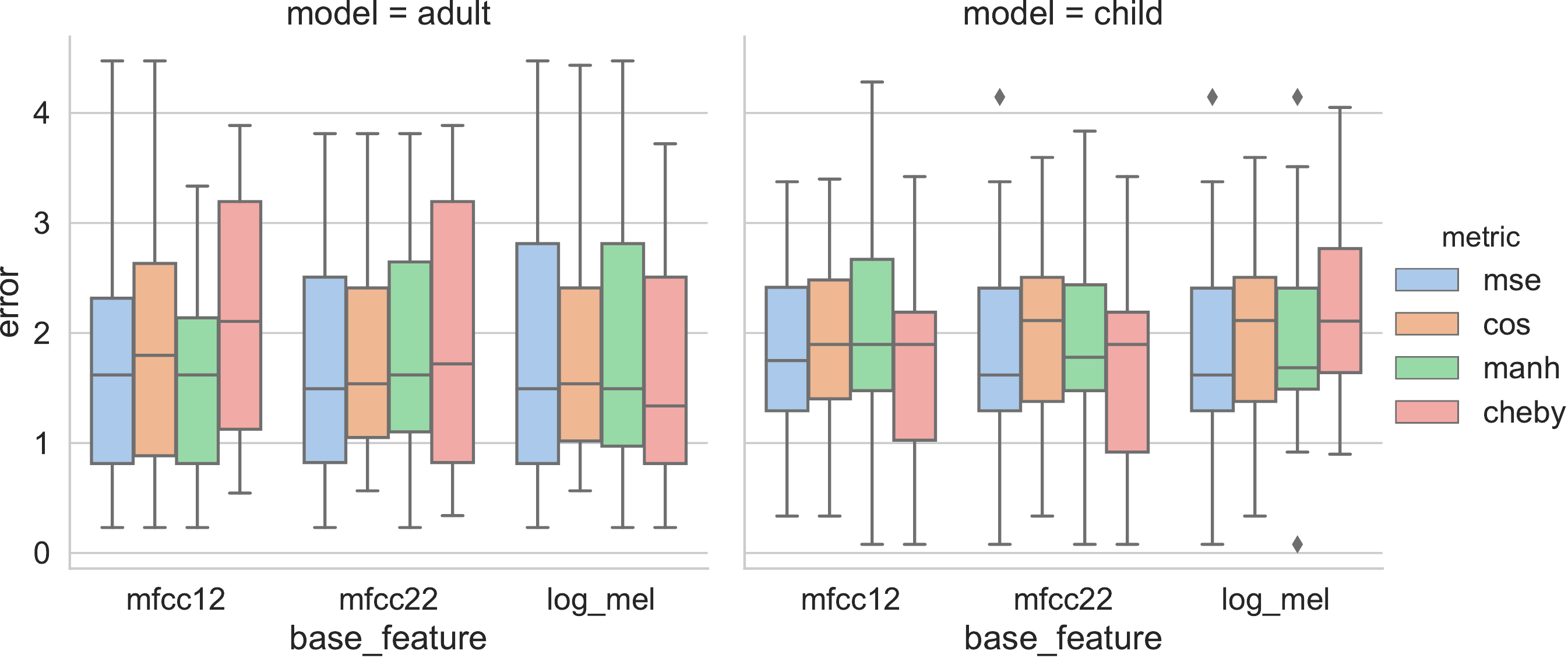}
  \caption{Impact of the different metrics on formant error for the base features without HF emphasis or normalisation.}
  \label{fig:metrics}
\vspace{-10pt}
\end{figure}

\begin{figure}[]
  \centering
  \includegraphics[width=\linewidth]{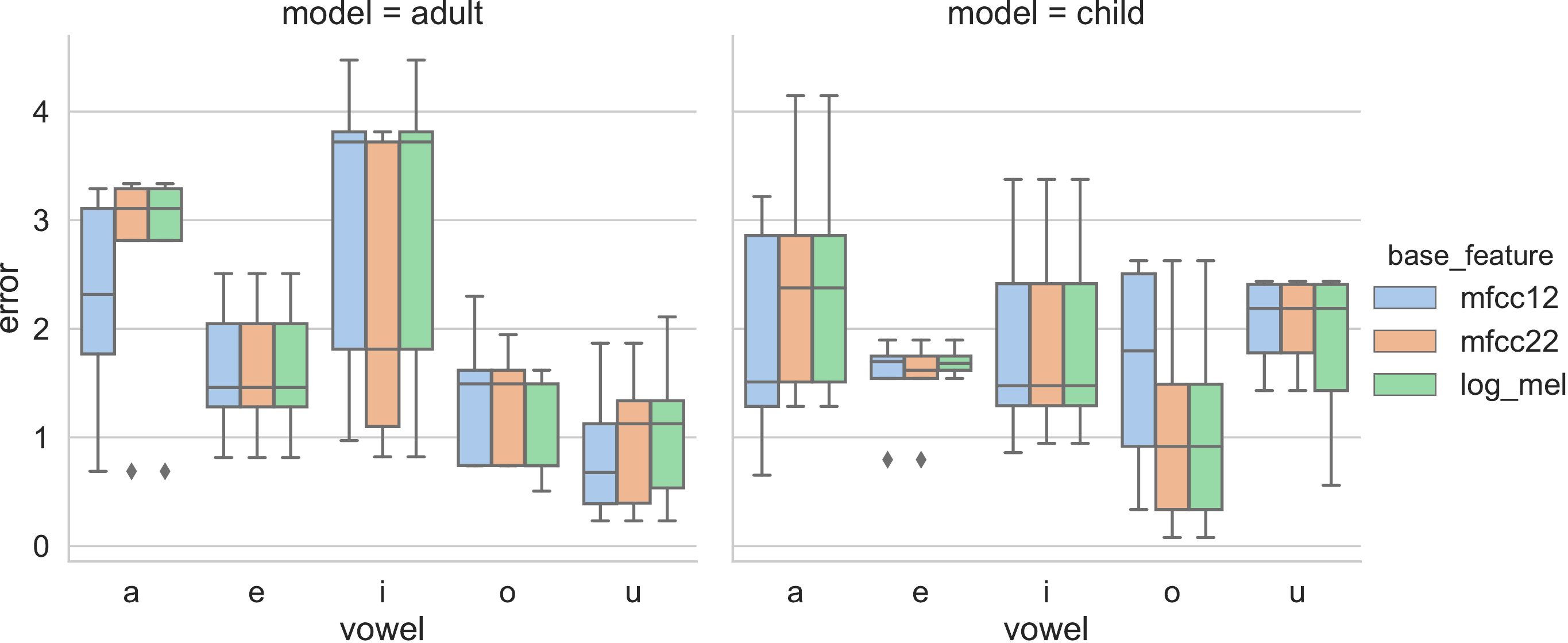}
  \caption{Impact of the different base features on formant error.}
  \label{fig:features}
\vspace{-10pt}
\end{figure}

\begin{figure*}[]
  \centering
  \includegraphics[width=.9\linewidth]{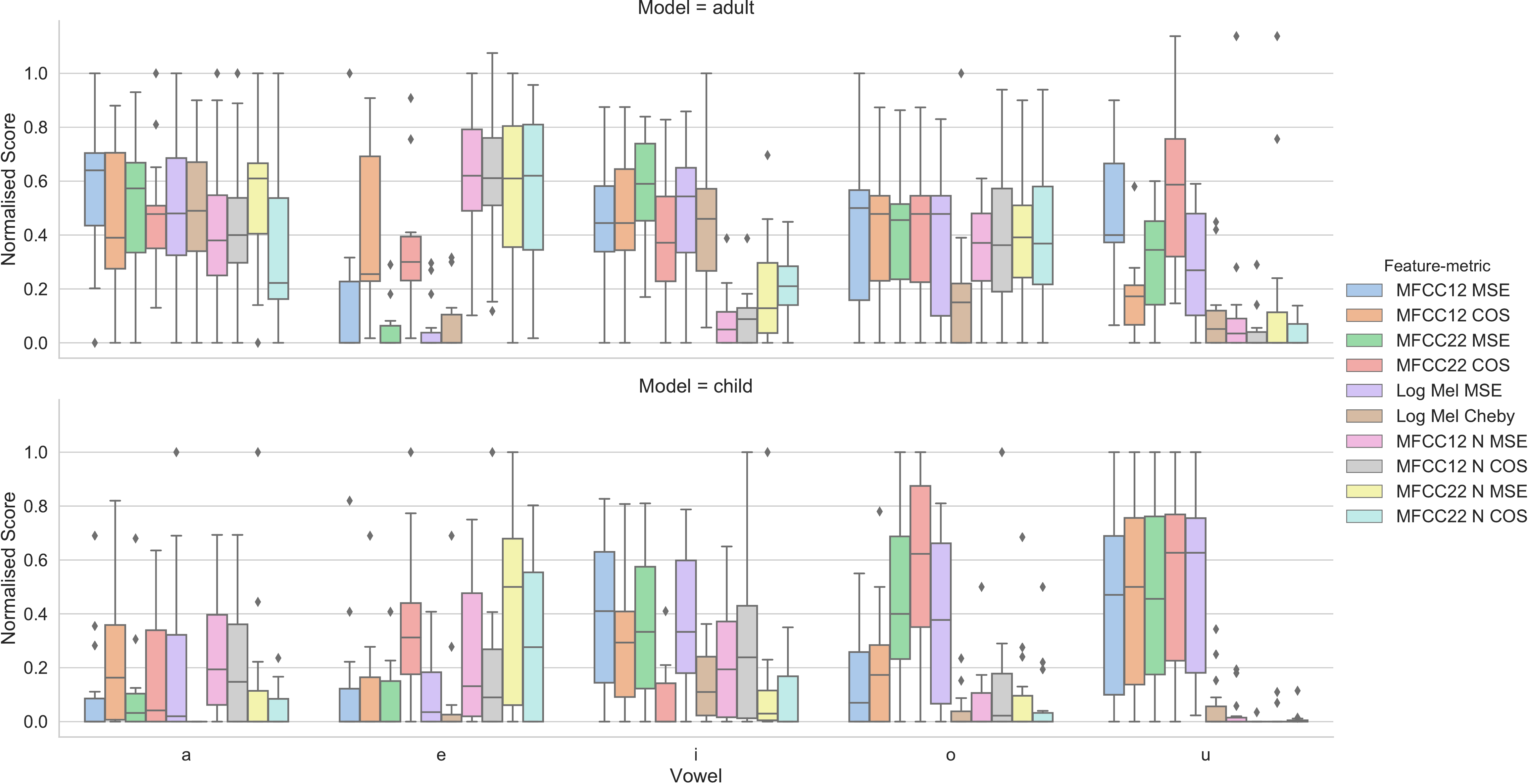}
  \caption{Normalised scores from the listening test.}
  \label{fig:listening}
\vspace{-10pt}
\end{figure*}
\begin{figure}[]
  \centering
  \includegraphics[width=.49\linewidth]{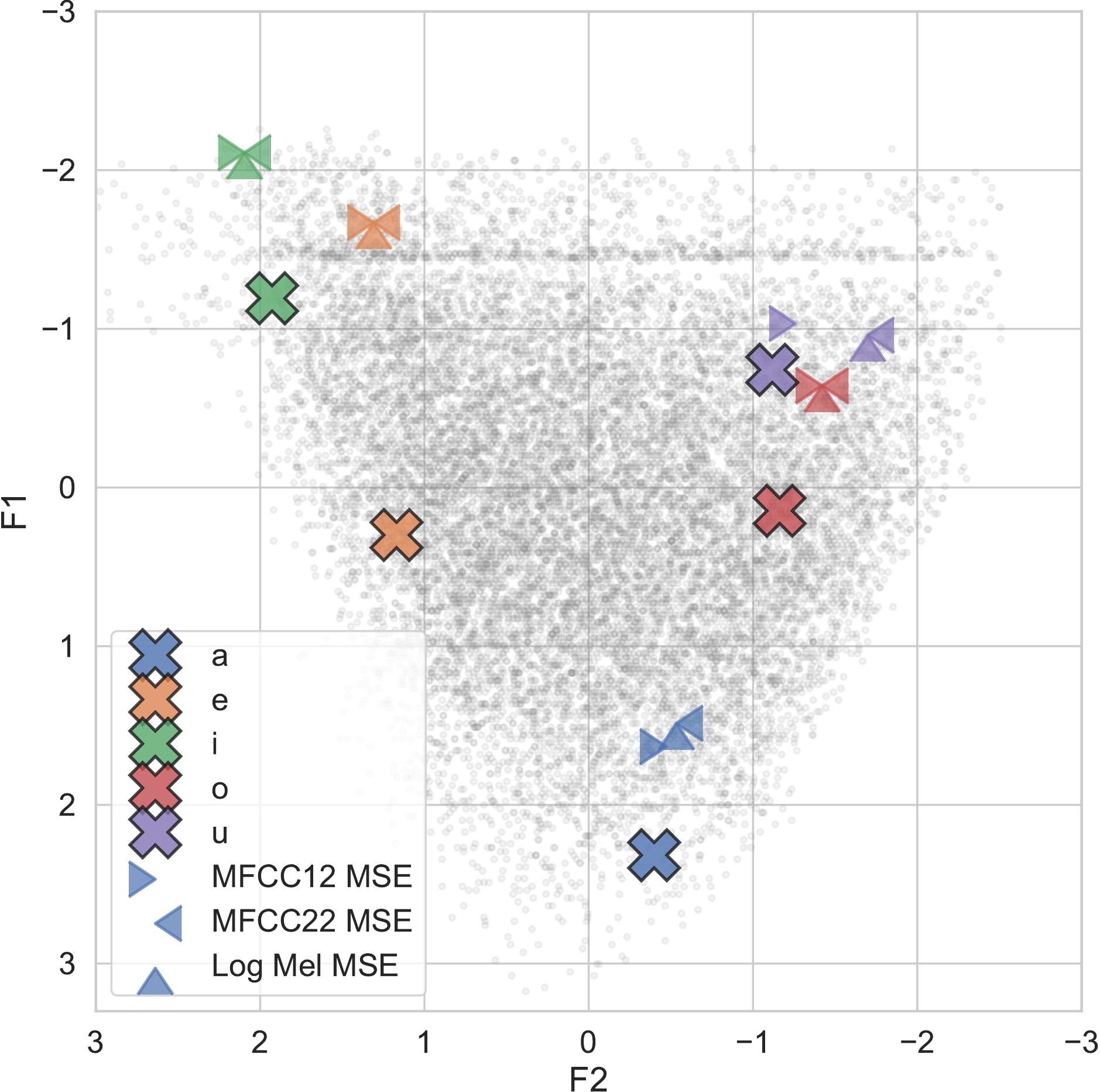}
  \includegraphics[width=.49\linewidth]{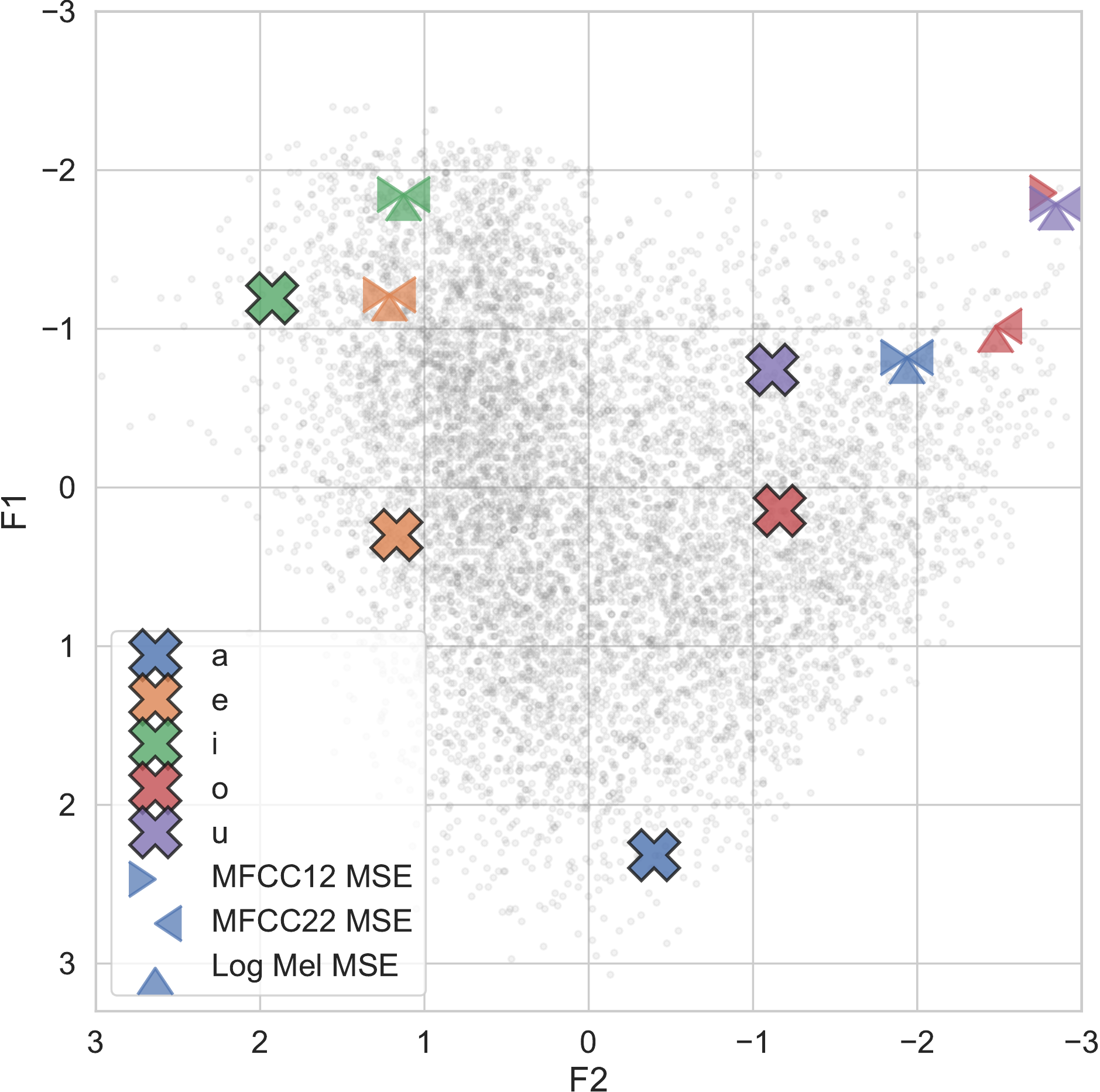}
  \caption{The formants of the optimised base features without normalisation used in the listening test for the adult (left) and child model (right).}
  \label{fig:listening_formants}
\vspace{-15pt}
\end{figure}

\section{Results}
\subsection{Formant space analysis}
\textbf{Impact of high frequency emphasis.}
The obtained formant error when using HF emphasis aggregated across the vowels, metrics, normalisation, and grouped by base feature for each model 
is shown in Fig.~\ref{fig:hfemph}.
We can see 
that the use of HF emphasis on average increases the error as measured by the distance to the target in the normalised F1-F2 space.

\noindent\textbf{Impact of normalisation.}
The formant error results do not reveal a clear cut impact of normalisation in the optimisation task.
Instead we investigate the error surface projections of MFCC12 MSE for /e/ and /u/ for the adult and child models shown in
Fig.~\ref{fig:surf_u}.
%
We can see that the impact of normalisation is more pronounced for /u/.
Indeed, while it only leads to a loss of the pronounced minimum, for the child model the effects of normalisation are severe, shifting the global minimum to a different formant location altogether.

%

\noindent\textbf{Impact of the metrics.}
The averaged impact of the metrics for all the vowels for the base features without HF emphasis and normalisation is shown in Fig.~\ref{fig:metrics}.
We can see that although their performance is close, MSE offers smaller error on average.

\noindent\textbf{Impact of the features.}
Fig.~\ref{fig:features} shows the averaged impact of the base features without HF emphasis and normalisation for the different vowels.
We can see that the different base features work consistently across the vowels and the two models.
There are cases where MFCC12 work better (adult /a/ and /u/ and child /a/), but also worse (child /o/).
This can be explained by the similarity of their error surfaces, as seen in Fig.~\ref{fig:surf_u}.

\subsection{Listening tests}
The overall results of the listening tests are shown in Fig.~\ref{fig:listening}.
We can see that the scores between raters are mostly consistent.
A stronger indicator is that there are feature-metric pairs that clearly resulted with a phonetically erroneous synthesis.
It is also indicative that there is a strong and inconsistent impact of normalisation on the different vowels.
Specifically, normalisation seems to systematically improve performance for /e/, while impairing it for /u/ for both models.
This phenomenon can be readily explained by the error surface projections for these two vowels, shown in Fig.~\ref{fig:surf_u}.
We can indeed see that normalisation shifts the global minimum of the error closer to the formant target /e/ and away from /u/.

If we focus on the scores obtained by the base features without normalisation and with the MSE metric, we can see that the rater scores are consistent for the different vowels, with some exceptions, which is in line with our observations of the formant error and their error surface similarity.
In fact, we can see in Fig.~\ref{fig:listening_formants} that most of these have selected the same synthesised vowel.
Moreover, the relative distance in formant space correlates well with the perceptual scores, i.e. the low scores for /e/ in both models and /a/ in the child model, as well as the worse result obtained for MFCC12 for the child model /o/, and its improved score for the adult /a/ and /u/.

If we examine the selected formant position for adult /e/ and compare it to the error surface shown in Fig.~\ref{fig:surf_u}, we can see that it does not coincide with the expected global minimum.
This is due to the variance of the binned errors around the calculated mean not shown here because of space limitations.


\section{Conclusions}

While formant error does not tell the whole story when it comes to the acoustic realisation of vowels, our findings show that normalised formant distance correlates well with perceptual scores of vowel quality.
We have also shown that the projection of the error surface in the normalised F1-F2 space can serve to evaluate feature-metric pairs and predict their perceptual performance for the optimisation of vocal tract parameters in simulations of vocal learning.
Moreover, these projections show wrong our intuition that there is a straightforward correspondence between error optimisation in the feature space and minimisation of formant error.

From the evaluated feature-metric pairs we have demonstrated similarity in the formant space error surfaces, formant errors and perceptual scores between the MFCC12, MFCC22 and Log Mel base features.
None of them has demonstrated superiority in the task of vowel production optimisation.
The performance of the different metrics is also similar, with MSE giving slightly better average results.
High frequency emphasis has shown to increase formant error
and should not be used for the task of vowel learning.
However, it might have a positive impact on consonant learning.
Finally, normalisation has been shown to have a contradicting and severe impact on the error surface dynamics – improving it for some vowels and degrading it for others.

\section{Acknowledgements}

This work has been funded by the Leverhulme Trust Research Project Grant RPG-2019-241: ``High quality simulation of early vocal learning''.
Formant analysis of target speaker was funded by the National Science Centre of Poland 2017/25/B/HS2/00760.


\bibliographystyle{IEEEtran}
\balance

\bibliography{refs}


\end{document}